\documentstyle[11pt]{article}

\def\fnote#1#2{\begingroup\def\thefootnote{#1}\footnote{#2}\addtocounter
{footnote}{-1}\endgroup}

\begin{document}

\hfill{UTTG-01-05}

\vspace{36pt}

\begin{center}
{\large {\bf {Quantum Contributions to Cosmological Correlations}}}

\vspace{36pt}
Steven Weinberg\fnote{*}{Electronic address:
weinberg@physics.utexas.edu}\\
{\em Theory Group, Department of Physics, University of
Texas\\
Austin, TX, 78712}

\vspace{30pt}

\noindent
{\bf Abstract}
\end{center}
\noindent
The ``in-in'' formalism is reviewed and extended, and applied to the calculation of higher-order Gaussian and non-Gaussian correlations in cosmology.  Previous calculations of these correlations amounted to the evaluation of tree graphs in the in-in formalism; here we also consider loop graphs.  It turns out that for some though not all theories, the contributions of loop graphs as well as tree graphs depend only on the behavior of the inflaton potential near the time of horizon exit.  A sample one-loop calculation is presented.

 \vfill

\pagebreak
\setcounter{page}{1}

\begin{center}
{\bf I. INTRODUCTION}
\end{center}

The departures from cosmological homogeneity and isotropy observed in the cosmic microwave background and large scale structure are small, so it is natural that they should be dominated by a Gaussian probability distribution, with bilinear averages given by the terms in the Lagrangian that are quadratic in perturbations.  Nevertheless, there is growing interest in the possibility of observing non-Gaussian terms in various correlation functions,$^1$ such as an expectation value of a product of three temperature fluctuations.  It is also important to understand the higher-order corrections to bilinear correlation functions, which appear in Gaussian correlations.

Until now, higher-order cosmological correlations have been calculated by solving the classical field equations beyond the linear approximation.  As will be shown in the Appendix, this is equivalent to calculating sums of tree graphs, though in a formalism different from the familiar Feynman graph formalism.  For instance, Maldacena$^2$ has calculated the non-Gaussian average of a product of three scalar and/or gravitational fields to first order in their interactions, which amounts to calculating a tree graph consisting of a single vertex with 3 attached gravitational and/or scalar field lines.  

This paper will discuss how calculations of cosmological correlations can be carried to arbitrary orders of perturbation theory, including the quantum effects represented by loop graphs.  So far, loop corrections to correlation functions appear to be much too small ever to be observed.  The present work is motivated by the opinion that we ought to understand what our theories entail, even where in practice its predictions cannot be verified experimentally, just as field theorists in the 1940s and 1950s took pains to understand quantum electrodynamics to all orders of perturbation theory, even though it was only possible to verify results in the first few orders.

There is a particular question that will concern us.  In the familiar calculations of lowest-order Gaussian correlations, and also in Maldacena's tree-graph calculation of non-Gaussian correlations, the results depended only on the behavior of the unperturbed inflaton field near the time of horizon exit.  Is the same true for loop graphs?  If so, it will be possible to calculated the loop contributions with some confidence, but we can learn 
little new from such calculations.  On the other hand, if the contribution of loop graphs depends on the whole history of the unperturbed inflaton field, then calculations become much more difficult, but potentially more revealing.  In this case, it might even be that the loop contributions are much larger than otherwise expected.  

The appropriate formalism for dealing with this sort of problem is the ``in-in'' formalism originally due to Schwinger.$^3$  Schwinger's presentation is somewhat opaque, so this formalism is outlined (and extended) in an Appendix.  In section II  we summarize  those aspects of this formalism that are needed for our present purposes.   Section III introduces a class of theories to serve as a basis of discussion, with a single inflaton field, plus any number of additional massless scalar fields with only gravitational interactions and vanishing expectation values.   In Section IV we prove a general theorem about the late time behavior of cosmological correlations at fixed internal as well as external wave numbers.  Section V  introduces a class of unrealistic theories to illustrate the problems raised by the integration over internal wave numbers, and how these problems may be circumvented.  In Section VI we return to the theories introduced in Section III, and we show that the conditions of the theorem proved in Section IV are satisfied for these theories.  This means that,  to all orders of perturbation theory, if ultraviolet divergences cancel in the integrals over internal wave numbers, then cosmological correlations do indeed depend only on the behavior of the unperturbed inflation field near the time of horizon exit in the cases studied.  We can also find other theories in which this result does not apply, as for instance by giving the additional scalar fields a self-interaction. Section VII presents a sample one-loop calculation of a cosmological correlation.

\begin{center}
{\bf II.  THE ``IN-IN'' FORMALISM IN COSMOLOGY}
\end{center}

The problem of calculating cosmological correlation functions differs from the more familiar problems encountered in quantum field theory in at least three respects:
\begin{itemize}
\item We are not interested here in the calculation of S-matrix elements, but rather in evaluating expectation values of products of fields at a fixed time.
\item Conditions are not imposed on the fields  at both very early and very late times, as in the calculation of S-matrix elements, but only at very early times, when the wavelength is deep inside the horizon and according to the Equivalence Principle the interaction picture fields should have the same form (when expressed in terms of metric rather than co-moving coordinates) as in Minkowski spacetime.
\item Although the Hamiltonian $H$ that generates the time dependence of the various quantum fields is constant in time, the time-dependence of the fluctuations in these fields are governed by a fluctuation Hamiltonian $\tilde{H}$ with an explicit time dependence, which as shown in the Appendix is constructed by expanding $H$  around the unperturbed solution of the field equation, and discarding the terms of first order in the perturbations to the fields and their canonical conjugates.
\end{itemize}

Given a fluctuation Hamiltonian $\tilde{H}$, we want to use it to calculate expectation values of some product $Q(t)$ of field operators, all at the same time $t$ but generally with different space arguments.  As discussed in the Appendix, the 
prescription of the ``in-in'' formalism is that 
\begin{equation} 
\langle Q(t)\rangle=\left\langle\left[\bar{T}\exp\left(i\int_{-\infty}^t H_I(t)\,dt\right)\right]\,Q^I(t)\,\left[T\exp\left(-i\int_{-\infty}^t H_I(t)\,dt\right)\right]\right\rangle\;,
\end{equation} 
Here $T$ denotes a time-ordered product; $\bar{T}$ is an anti-time-ordered product; $Q^I$ is the product $Q$ in the interaction picture (with time-dependence generated by the part of $\tilde{H}$ that is quadratic in fluctuations); and $H_I$ is the interaction part of $\tilde{H}$ in the interaction picture. (This result is different from that originally given by Maldacena$^2$ and other authors$^4$, who left out the time-ordering and anti-time-ordering, perhaps  through a typographical error.  
However, this makes no difference to first order in the interaction, which is the approximation used by these authors  in their calculations.)  We are here taking the time $t_0$ at which the fluctuations are supposed to behave like free fields as $t_0=-\infty$, which is appropriate for cosmology because at very early times the fluctuation wavelengths are deep inside the horizon.

Eq.~(1) leads to a fairly complicated diagrammatic formalism, described in the Appendix.  Unfortunately this formalism obscures crucial cancellations that occur between different diagrams.  For our present purposes, it is more convenient to use a formula equivalent to Eq.~(1):
\begin{eqnarray}
\langle Q(t)\rangle &=&\sum_{N=0}^\infty i^N\, \int_{-\infty}^t dt_N \int_{-\infty}^{t_N} dt_{N-1} \cdots \int_{-\infty}^{t_2} dt_1 \nonumber\\&&\times\left\langle \Big[H_I(t_1),\Big[H_I(t_2),\cdots \Big[H_I(t_N),Q^I(t)\Big]\cdots\Big]\Big]\right\rangle\;,
\end{eqnarray}
(with the $N=0$ term   understood to be just $\langle Q^I(t)\rangle $).  This can easily be derived from Eq.~(1) by mathematical induction.  Obviously Eqs.~(1) and (2) give the same results to zeroth and first order in $H_I$.  If we assume that the right-hand sides of Eqs.~(1) and (2) are equal for arbitrary operators $Q$ up to order $N-1$ in $H_I$, then by differentiating these equations we easily see that the time derivatives of the right-hand sides are equal up to order $N$.  Eqs.~(1) and (2) also give the same results for $t\rightarrow -\infty$ to all orders, so they give the same results for arbitrary $t$ to order $N$.

\vspace{12pt}

\begin{center}
{\bf III. THEORIES OF INFLATION}
\end{center}

To make our discussion concrete, in this section we will take up a particular class of theories of inflation.  The reader who prefers to avoid details of specific theories can skip this section, and go on immediately to the general analysis of late-time behavior in the following section.

In this section we will consider theories of inflation with two kinds of  matter fields : a real scalar field $\varphi({\bf x},t)$ with a non-zero homogeneous expectation value $\bar{\varphi}(t)$ that rolls down a potential $V(\varphi)$, and any number of  real massless scalar fields $\sigma_n({\bf x},t)$, which have only minimal gravitational interactions, and are prevented by unbroken symmetries from acquiring vacuum expectation values.  The real field $\varphi$ serves as an inflaton whose energy density drives inflation, while the  $\sigma_n$ are a stand-in for the large number of species of matter fields that will dominate the effects of loop graphs on the  correlations of the inflaton field.\fnote{**}{Standard counting arguments show that in these theories the number of factors of $8\pi G $ in any graph equals the number of loops of any kind,  plus a fixed number that depends only on which correlation function is being calculated.  Matter loops are numerically more important than loops containing graviton or inflaton lines, because they carry an additional factor equal to the number of types of matter fields.}

We follow Maldacena,$^2$ adopting a gauge in which there are no fluctuations in the inflaton field, so that $\varphi({\bf x},t)=\bar{\varphi}(t)$, and in which the spatial part of the metric takes the form\fnote{***}{I am adopting Maldacena's notation, but the quantity he calls $\zeta$ is more usually called ${\cal R}$.  To first order in fields, the quantity usually called $\zeta$ is defined as $-\Psi-H\delta\rho/\dot{\bar{\rho}}$, while the quantity usually called ${\cal R}$  is defined as $-\Psi+H\delta u$.  (Here the contribution of scalar modes to $g_{ij}$ is written in general gauges as $-2a^2(\Psi\delta_{ij}+\partial^2\Psi'/\partial x^i\partial x^j)$, while  $\delta\rho$ and $\bar{\rho}$ are the perturbation to the total energy density and its unperturbed value, while  $\delta u$ is the perturbed velocity potential, which for a single inflaton field is $\delta u=-\delta\varphi/\dot{\bar{\varphi}}$.)  In the gauge used by Maldacena and in the present  paper $\delta u=\Psi'=0$, so since $\zeta$ is defined here as $\Psi$ to first order in fields, it corresponds to the quantity usually called ${\cal R}$.  Outside the horizon ${\cal R}$ and $\zeta$ are the same.}

\begin{equation}
g_{ij}=a^2e^{2\zeta}[\exp{\gamma}]_{ij} ,~~~~~\gamma_{ii}=0,~~~~~\partial_i\gamma_{ij}=0\;.
\end{equation}
where $a(t)$ is the Robertson--Walker scale factor, $\gamma_{ij}({\bf x},t)$ is a gravitational wave amplitude, and $\zeta({\bf x},t)$ is a scalar whose  characteristic feature  is that it is conserved outside the horizon,$^5$ that is, for physical wave numbers that are small compared with the expansion rate.  The same is true of $\gamma_{ij}$.

The other components of the metric are given in the Arnowitt--Deser--Misner (ADM) formalism$^6$ by
\begin{equation}
g_{00}=-N^2+g_{ij}N^iN^j\;,~~~~~~~g_{i0}=g_{ij}N^j\;,
\end{equation}
where $N$ and $N^i$ are auxiliary fields, whose time-derivatives do not appear in the action.
The Lagrangian density in this gauge (with $8\pi G\equiv 1$) is
\begin{eqnarray}
&&{\cal L}=\frac{a^3}{2}e^{3\zeta}\Bigg[NR^{(3)}-2NV(\bar{\varphi})+N^{-1}\Big(E^j{}_iE^i{}_j-(E^i{}_i)^2\Big)+N^{-1}\dot{\bar{\varphi}}^2\nonumber\\&&~~~~~~+N^{-1}\sum_n\Big(\dot{\sigma}_n-N^i\partial_i\sigma_n\Big)^2-Na^{-2}e^{-2\zeta}[\exp{(-\gamma)}]^{ij}\sum_n\partial_i\sigma_n\partial_j\sigma_n\Bigg]\;,\nonumber\\&&{}
\end{eqnarray}
where 
\begin{equation}
E_{ij}\equiv \frac{1}{2}\Big(\dot{g}_{ij}-\nabla_iN_j-\nabla_jN_i\Big)\;,
\end{equation}
and bars denote unperturbed quantities.
All spatial indices $i$, $j$, etc. are lowered  and raised  with the matrix $g_{ij}$ and its reciprocal; $\nabla_i$ is the three-dimensional covariant derivative calculated with this three-metric; and $R^{(3)}$ is the curvature scalar calculated with this three-metric:
$$
R^{(3)}=a^{-2}e^{-2\zeta}\Big[e^{-\gamma}\Big]^{ij}R^{(3)}_{ij}.
$$
The auxiliary fields $N$ and $N^i$ are to be found by requiring that the action is stationary in these variables.  This gives the constraint equations:
\begin{equation}
\nabla_i\left[N^{-1}\Big(E^i{}_j-\delta^i{}_jE^k{}_k\Big)\right]=N^{-1}\sum_n\partial_j\sigma_n\Big(\dot{\sigma}_n-N^i\partial_i\sigma_n\Big)\;,
\end{equation}
\begin{eqnarray}
&& N^2\left[R^{(3)}-2V-a^{-2}e^{-2\zeta}[\exp(-\gamma)]^{ij}\sum_n\partial_i\sigma_n
\partial_j\sigma_n\right]=E^i{}_jE^j{}_i-\left(E^i{}_i\right)^2\nonumber\\&&~~~~~~~~~~~~+\dot{\bar{\varphi}}^2
+\sum_n\Big(\dot{\sigma}_n-N^i\partial_i\sigma_n\Big)^2
\end{eqnarray}
For instance, to first order in fields (including field derivatives) the auxiliary fields are
the same as in the case of no additional matter fields$^2$
\begin{equation}
N=1+\dot{\zeta}/H\;,~~~N^i=-\frac{1}{a^2H}\partial_i\zeta+\epsilon\partial_i\nabla^{-2}\dot{\zeta}\;,
\end{equation}
where 
\begin{equation} 
\epsilon\equiv -\frac{\dot{H}}{H^2}=\frac{\dot{\bar{\varphi}}^2}{2H^2}\;,~~~~~H\equiv \frac{\dot{a}}{a}\;
\end{equation} 

The fields in the interaction picture
satisfy free-field equations.  For $\zeta$ we have the Mukhanov equation:$^7$
\begin{equation} 
\frac{\partial^2\zeta}{\partial t^2} +\left[\frac{d \ln(a^3\epsilon)}{dt}\right]\frac{\partial\zeta }{\partial t}-a^{-2}\nabla^2\zeta=0\;,
\end{equation} 
The field equation for gravitational waves is
\begin{equation} 
\frac{\partial^2\gamma_{ij}}{\partial t^2} +3H\frac{\partial\gamma_{ij} }{\partial t}-a^{-2}\nabla^2\gamma_{ij}=0\;,
\end{equation} 
and for the matter fields
\begin{equation}
\frac{\partial^2\sigma_n}{\partial t^2}+3H\frac{\partial\sigma_n}{\partial t}-a^{-2}\nabla^2\sigma_n=0\;.
\end{equation}
The fields in the interaction picture are then
\begin{equation} 
\zeta({\bf x},t)=\int d^3q\left[e^{i{\bf q}\cdot{\bf x}}\alpha({\bf q})\zeta_q(t)+
e^{-i{\bf q}\cdot{\bf x}}\alpha^*({\bf q})\zeta^*_q(t)\right]\;,
\end{equation} 
\begin{equation} 
\gamma_{ij}({\bf x},t)=\int d^3q\sum_\lambda\left[e^{i{\bf q}\cdot{\bf x}}e_{ij}(\hat{q},\lambda)\alpha({\bf q},\lambda)\gamma_q(t)+
e^{-i{\bf q}\cdot{\bf x}}e^*_{ij}(\hat{q},\lambda)\alpha^*({\bf q},\lambda)\gamma^*_q(t)\right]\;,
\end{equation}
\begin{equation} 
\sigma_n({\bf x},t)=\int d^3q\left[e^{i{\bf q}\cdot{\bf x}}\alpha({\bf q},n)\sigma_q(t)+
e^{-i{\bf q}\cdot{\bf x}}\alpha^*({\bf q},n)\sigma^*_q(t)\right]\;,
\end{equation}
where $\lambda=\pm 2$ is a helicity index and $e_{ij}(\hat{q},\lambda)$ is a polarization tensor, while $\alpha({\bf q})$,  $\alpha({\bf q},\lambda)$, and $\alpha({\bf q},n)$ are conventionally normalized  annihilation operators, satisfying the usual commutation relations
\begin{equation} 
\Big[\alpha({\bf q})\,,\,\alpha^*({\bf q}')\Big]=\delta^3\Big({\bf q}-{\bf q}'\Big)\;,~~~~~~
\Big[\alpha({\bf q})\,,\,\alpha({\bf q}')\Big]=0\;.
\end{equation}
\begin{equation} 
\Big[\alpha({\bf q},\lambda)\,,\,\alpha^*({\bf q}',\lambda')\Big]=\delta_{\lambda\lambda'}\delta^3\Big({\bf q}-{\bf q}'\Big)\;,~~~~~~
\Big[\alpha({\bf q},\lambda)\,,\,\alpha({\bf q}',\lambda')\Big]=0\;,
\end{equation} 
and 
\begin{equation}
\Big[\alpha({\bf q},n)\,,\,\alpha^*({\bf q}',n')\Big]=\delta_{nn'}\delta^3\Big({\bf q}-{\bf q}'\Big)\;,~~~~~~
\Big[\alpha({\bf q},n)\,,\,\alpha({\bf q}',n')\Big]=0\;,
\end{equation}
Also, $\zeta_q(t)$, $\gamma_q(t)$, and $\sigma_q(t)$ are suitably normalized 
positive-frequency solutions of Eqs.~(11)--(13), with $\nabla^2$ replaced with $-q^2$.
They satisfy initial conditions, designed to make $-\zeta\dot{\bar{\varphi}}/H$, $\gamma_{ij}/\sqrt{16\pi G}$, and $\sigma_n$ behave like conventionally normalized free fields at $t\rightarrow-\infty$:\fnote{\dagger}{In Newtonian gauge the quantity $-\zeta({\bf x},t)\dot{\bar{\varphi}}(t)/H(t)$ approaches the  inflaton field fluctuation $\delta\varphi(t)$ for $t\rightarrow -\infty$.} 
\begin{eqnarray}
&&-\frac{\dot{\bar{\varphi}}(t)\zeta_q(t)}{H(t)}\rightarrow \frac{\gamma_q(t)}{\sqrt{16\pi G}}\rightarrow \sigma_q(t)\nonumber\\&&~~~~~~\rightarrow \frac{1}{(2\pi)^{3/2}\sqrt{2q}\,a(t)}\exp\left(iq\int^\infty_t\frac{dt'}{a(t')}\right)\;.
\end{eqnarray}

\vspace{12pt}

\begin{center}
{\bf IV. LATE TIME BEHAVIOR}
\end{center}

The  question to be addressed in this section is whether the time integrals in Eqs.~(1) and (2) are  dominated by times near horizon exit for general graphs.  
This question is more complicated for loop graphs than for tree graphs, such as that considered by Maldacena,  because for loops there are two different kinds of wave number: the fixed wave numbers $q$ associated with external lines, and the internal wave numbers $p$ circulating in loops, over which we must integrate.  It is only if the integrals over internal wave numbers $p$ are dominated by values of order $p\approx q$ that we can speak of a definite time of horizon exit, when $q/a\approx p/a\approx H$.  In this section we will integrate first over the time arguments in Eq.~(2), holding the internal wave numbers at fixed values, and return at the end of this section to the problems raised by the necessity of then integrating over the $p\,$s.

There is never any problem with the convergence of the time integrals at very early times; all fluctuations oscillate very rapidly for $q/a\gg H$ and $p/a\gg H$, suppressing the contribution of early times to the time integrals in Eq.~(2).  To see what happens for late times, when $q/a\ll H$ and $p/a\ll H$, we need to count the powers of $a$ in the contribution of late times in general loop as well as tree graphs.  

For this purpose, we need to consider the behavior of the coefficient functions appearing in the Fourier decompositions (14)--(16) of the fields in the interaction picture.  In order to implement dimensional regularization, we will consider these coefficient functions in $2\nu$ space dimensions, returning later to the limit $2\nu\rightarrow 3$.  The coefficient functions then obey  differential equations obtained by replacing the space dimensionality 3 in Eqs.~(11)--(13) with $2\nu$, as well as replacing the Laplacian with $-q^2$:
\begin{equation} 
\frac{d^2\zeta_q(t)}{d t^2} +\left[\frac{d \ln\Big(a^{2\nu}(t)\epsilon(t)\Big)}{dt}\right]\frac{d\zeta_q(t) }{d t}+\frac{q^2}{a^2(t)}\zeta_q(t)=0\;,
\end{equation} 
\begin{equation} 
\frac{d^2\gamma_q(t)}{d t^2} +2\nu H(t)\frac{d\gamma_q(t) }{d t}+\frac{q^2}{a^2(t)}\gamma_q(t)=0\;,
\end{equation} 
\begin{equation}
\frac{d^2\sigma_q(t)}{d t^2}+2\nu H(t)\frac{d\sigma_q(t)}{d t}+\frac{q^2}{a^2}\sigma_q(t)=0\;.
\end{equation}
At late times,   when $q/a\ll H$, the solutions can be written as asymptotic expansions in inverse powers of $a$:\fnote{\dagger\dagger}{By $t= \infty$ in the limits of these integrals and elsewhere in this paper, we  mean  a time still during inflation, but  sufficiently late  so that $a(t)$ is many e-foldings larger than its value when $q/a$ falls below $H$.}
\begin{eqnarray}
&&\zeta_q(t)\rightarrow \zeta_q^o\left[1+\int_t^\infty\frac{q^2\,dt'}{a^{2\nu}(t')\epsilon(t')}\int_{-\infty}^{t'} a^{2\nu-2}(t'')\,\epsilon(t'')\,dt''+\dots\right]
\nonumber\\&& ~~~~~~+C_q\Bigg[\int_t^\infty\frac{dt'}{a^{2\nu}(t')\epsilon(t')}\nonumber\\&&+q^2\int_t^\infty \frac{dt'}{a^{2\nu}(t')\epsilon(t')}\int_{-\infty}^{t'}a^{2\nu-2}(t'')\,\epsilon(t'')\,dt''\int_{t''}^\infty\frac{dt'''}{a^{2\nu}(t''')\epsilon(t''')}+\dots\Bigg]\\
&&\gamma_q(t)\rightarrow \gamma_q^o\left[1+\int_t^\infty\frac{q^2\,dt'}{a^{2\nu}(t')}\int_{-\infty}^{t'} a^{2\nu-2}(t'')\,dt''+\dots\right]\nonumber\\&&+D_q\left[\int_t^\infty\frac{dt'}{a^{2\nu}(t')}+q^2\int_t^\infty
\frac{dt'}{a^{2\nu}(t')}\int_{-\infty}^{t'}a(t'')\,dt''\int_{t''}^\infty \frac{dt'''}{a^{2\nu}(t''')}+\dots\right]
\\
&&\sigma_{q}(t)\rightarrow \sigma_q^o\left[1+\int_t^\infty\frac{q^2\,dt'}{a^{2\nu}(t')}\int_{-\infty}^{t'} a^{2\nu-2}(t'')\,dt''+\dots\right]\nonumber\\&&+E_q\left[\int_t^\infty\frac{dt'}{a^{2\nu}(t')}+q^2\int_t^\infty
\frac{dt'}{a^{2\nu}(t')}\int_{-\infty}^{t'}a^{2\nu-2}(t'')\,dt''\int_{t''}^\infty \frac{dt''}{a^{2\nu}(t'')}+\dots\right]
\nonumber\\&&{}
\end{eqnarray}
where $\zeta^o_q$, $\gamma_q^0$, and $\sigma_q^o$ are the limiting values of 
$\zeta_q(t)$, $\gamma_q(t)$, and $\sigma_{q}(t)$ (the ``$o$'' superscript stands for ``outside the horizon") and $C_q$, $D_q$, and $E_q$ are additional constants.  In any kind of inflation with sufficient expansion, the Robertson-Walker scale factor $a$ will grow much faster than $H$ or $\epsilon$ can change, and Eqs.~(24)--(26) thus show that (at least for $2\nu\geq 2$) the time derivatives of $\zeta_q$, $\gamma_q$, and $\sigma_q$ all vanish for $q/a\ll H$ like $1/a^2$.  

If an interaction involves enough factors of $\dot{\zeta}$, $\dot{\gamma}_{ij}$, and/or $\dot{\sigma}_{n}$ so that these $1/a^2$ factors and any $1/a^2$ factors from the  contraction of space indices more than compensate for the $a^{2\nu}$ factor in the interaction  from the square root of the metric determinant, then the integral over the associated time coordinate will converge exponentially fast at late times as well as at early times, and therefore may be expected to be dominated by the era in which the wavelength leaves the horizon.
For instance, the extension of Eq.~(5) to $2\nu$ space dimensions gives the interaction between a $\zeta$ field and a pair of $\sigma$ fields  
\begin{eqnarray}
{\cal L}_{\zeta\sigma\sigma}&=&-\frac{a^{2\nu-2}}{2}\zeta\sum_n\partial_i\sigma_n\,\partial_i\sigma_n-\frac{a^{2\nu-2}}{2H}\dot{\zeta}\sum_n
\partial_i\sigma_n\,\partial_i\sigma_n
\nonumber\\&&+a^{2\nu-2}\partial_i\left(\frac{\zeta}{H}-\epsilon a^2\nabla^{-2}\dot{\zeta}\right)\sum_n \dot{\sigma}_n\partial_i\sigma_n
\nonumber\\&&
-\frac{a^{2\nu}}{2H}\dot{\zeta}\sum_n\dot{\sigma}_n^2+\frac{3a^{2\nu}}{2}\zeta \sum_n\dot{\sigma}_n^2\;.
\end{eqnarray}
(The $\zeta\sigma\sigma$ interaction Hamiltonian given by canonical quantization is just $-\int d^{2\nu}x\; {\cal L}_{\zeta\sigma\sigma}$, 
but this simple relation does not always apply.)  Counting a factor $a^{-2}$ for each $\dot{\zeta}$ or $\dot{\sigma}_n$, the terms in this interaction go as $a^{2\nu-2}$, $a^{2\nu-4}$, $a^{2\nu-4}$, $a^{2\nu-6}$, and $a^{2\nu-4}$, respectively.  All these terms are safe for $2\nu<4$, except for the first, which for $2\nu>2$ grows exponentially   at late times.   

Because of the commutators in Eq.~(2), the condition for a safe interaction is actually less stringent than that it should decay exponentially with time, and even a growing term that only involves fields rather than their time derivatives, like the first term in Eq.~(27), may not destroy the convergence of the time integrals.  We will now prove the following:

\vspace{12pt}

\noindent
{\bf Theorem}
The integrals over the time coordinates of interactions converge exponentially for $t\rightarrow \infty$, essentially as $\int^\infty dt/a^n(t)$ with $n>0$, provided that in $2\nu$ space dimensions, all interactions are of one or the other of two types:
\begin{itemize}
\item Safe interactions, that contain a number of factors of $a(t)$ (including  $-2$ factors of $a$ for each time derivative and the $2\nu$ factors of $a$ from $\sqrt{-{\rm Det}g}$) strictly less than $2\nu-2$, and
\item Dangerous interactions, which grow at late times no faster than $a^{2\nu-2}$, {\em and contain only fields, not time derivatives of fields}.
\end{itemize}
These conditions are evidently met by the interaction (27), irrespective of the value of $\nu$, and, as we shall see in Section VI, they are satisfied by all other interactions in the theories of Section III, but not in all theories.

Before proceeding to the proof, it should be noted that just as in Eq.~(27), the space dimensionality $2\nu$ enters in the interaction only in a factor $\sqrt{-{\rm Det}g}\propto a^{2\nu}$, so the question of whether or not a given theory satisfies the conditions of this theorem does not depend on the value of $2\nu$.  Thus this theorem has the corollary:

\vspace{12pt}

\noindent
{\bf Corollary}
The integrals over the time coordinates of interactions converge exponentially for $t\rightarrow \infty$, essentially as $\int^\infty dt/a^n(t)$ with $n>0$, provided that in 3 space dimensions all interactions are of one or the other of two types:
\begin{itemize}
\item Safe interactions, that contain a number of factors of $a(t)$ (including  $-2$ factors of $a$ for each time derivative and the 3 factors of $a$ from $\sqrt{-{\rm Det}g}$) strictly less than $+1$, and
\item Dangerous interactions, which grow at late times no faster than $a$, {\em and contain only fields, not time derivatives of fields}.
\end{itemize}

Here is the proof.  As already mentioned, the reason that dangerous interactions are not necessarily fatal has to do with how they enter into commutators in Eq.~(2). Because of the time-ordering in Eq.~(2), any failure of convergence of the time integrals for $t\rightarrow +\infty$   in $N$th-order perturbation theory must come from a region of the multi-time region of integration in which, for some $r$, the time arguments $t_r,\; t_{r+1},\;\dots t_N$, all go to infinity together.   We will therefore have to count the number of factors of $a(t_r),\,a(t_{r+1}),\;\dots a(t_N)$, treating them all as being of the same order of magnitude.  (This does not take proper account of factors of $\log a$, but as long as the integral over $t_r,\; t_{r+1},\;\dots t_N$ involves a negative total number of factors of $a$, it converges exponentially fast no matter how many factors of $\log a$ arise from subintegrations.)  Now, at least one of the fields or field time derivatives in each term in $H(t_s)$ with $r\leq s\leq N$ must appear in a commutator with one of the fields in some other $H_I(t_{s'})$ with $s<s'\leq N$.  So we need to consider the commutators of fields at times which may be unequal, but 
are both late.  In the sense described above, treating all $a(t_r),\,a(t_{r+1}),\;\dots a(t_N)$ as being of the same order of
 magnitude, if $a(t)$ increases more-or-less exponentially, then the 
commutator of two fields or a field and a field time-derivative goes as 
$a^{-2\nu}$, while the commutator of two field time-derivatives goes as 
$a^{-2\nu-2}$.

For instance, the  unequal-time commutators of the interaction-picture fields (14)--(16) are \begin{equation} 
\Big[\zeta({\bf x},t), \zeta({\bf x}',t')\Big]=\int d^{2\nu}p\; e^{i{\bf p}\cdot({\bf x}-{\bf x}')}
\Big(\zeta_p(t)\zeta^*_p(t')- \zeta_p(t')\zeta^*_p(t)\Big)\;,
\end{equation} 
\begin{equation} 
\Big[\gamma_{ij}({\bf x},t), \gamma_{kl}({\bf x}',t')\Big]=\int d^{2\nu}p\; e^{i{\bf p}\cdot({\bf x}-{\bf x}')}\,\Pi_{ijkl}(\hat{p})
\Big(\gamma_p(t)\gamma^*_p(t')- \gamma_p(t')\gamma^*_p(t)\Big)\;,
\end{equation} 
\begin{equation} 
\Big[\sigma_n({\bf x},t), \sigma_m({\bf x}',t')\Big]=\delta_{nm}\int d^{2\nu}p\; e^{i{\bf p}\cdot({\bf x}-{\bf x}')}
\Big(\sigma_p(t)\sigma^*_p(t')- \sigma_p(t')\sigma^*_p(t)\Big)\;,
\end{equation}
where $\Pi_{ijkl}(\hat{p})\equiv\sum_\lambda e_{ij}(\hat{p},\lambda) e_{kl}(\hat{p},\lambda)$.
The two asymptotic expansions given in Eqs.(21--(23) for each of the fields are both real aside from over-all factors, so neither by itself contributes to the commutators.  On the other hand, the constants $C_p\zeta^{o*}_p$, $D_p\gamma^{o*}_p$, and $E_p\sigma_p^{o*}$   are in general complex.  (For instance, in a strictly exponential expansion, inflation,  the phase of $C_p\zeta^{o*}_p $ is given by a factor $-e^{-i\nu \pi}$.)  The asymptotic expansions of the commutators at late times are therefore 
\begin{eqnarray} 
\Big[\zeta({\bf x}_1,t_1), \zeta({\bf x}_2,t_2)\Big]&\rightarrow& 2i\,\int d^{2\nu}p\; {\rm Im}\Big[ C_p\zeta_p^{o*} \Big]\;e^{i{\bf p}\cdot({\bf x}_1-{\bf x}_2)}\Bigg[\int_{t_1}^{t_2}\frac{dt'}{a^{2\nu}(t')\epsilon(t')}\nonumber
\\&& +p^2\int_{t_1}^{t_2}\frac{dt'}{a^{2\nu}(t')\epsilon(t')}\int_{-\infty}^{t'}a^{2\nu-2}(t'')\,\epsilon(t'')\,dt''\int_{t''}^\infty \frac{dt''}{a^{2\nu}(t'')\,\epsilon(t'')}\nonumber\\&&
+p^2\int_{t_1}^\infty \frac{dt'_1}{a^{2\nu}(t'_1)\,\epsilon(t'_1)}\int_{t_2}^\infty \frac{dt'_2}{a^{2\nu}(t'_2)\,\epsilon(t'_2)}\int_{t'_1}^{t'_2}a^{2\nu-2}(t'')\epsilon(t'')\,dt''+\dots\Bigg]
\;,\nonumber\\&&{}\\
\Big[\gamma_{ij}({\bf x}_1,t_1), \gamma_{kl}({\bf x}_2,t_2)\Big]&\rightarrow& 2i\,\int d^{2\nu}p\; \Pi_{ijkl}(\hat{p})\,{\rm Im} \Big[D_p\gamma_p^{o*}\Big]\;e^{i{\bf p}\cdot({\bf x}_1-{\bf x}_2)}
\Bigg[\int_{t_1}^{t_2}\frac{dt'}{a^{2\nu}(t')}\nonumber
\\&& +p^2\int_{t_1}^{t_2}\frac{dt'}{a^{2\nu}(t')}\int_{-\infty}^{t'}a^{2\nu-2}(t'')\,dt''\int_{t''}^\infty \frac{dt''}{a^{2\nu}(t'')}\nonumber\\&&
+p^2\int_{t_1}^\infty \frac{dt'_1}{a^{2\nu}(t'_1)}\int_{t_2}^\infty \frac{dt'_2}{a^{2\nu}(t'_2)}\int_{t'_1}^{t'_2}a^{2\nu-2}(t'')\,dt''+\dots\Bigg]\;,\nonumber\\&&{}\\ 
\Big[\sigma_n({\bf x}_1,t_1), \sigma_m({\bf x}_2,t_2)\Big]&\rightarrow& 2i\,\delta_{nm}\int d^{2\nu}p \;{\rm Im}\Big[E_p\sigma_p^{o*}\Big]\;e^{i{\bf p}\cdot({\bf x}_1-{\bf x}_2)}\Bigg[\int_{t_1}^{t_2}\frac{dt'}{a^{2\nu}(t')}\nonumber
\\&& +p^2\int_{t_1}^{t_2}\frac{dt'}{a^{2\nu}(t')}\int_{-\infty}^{t'}a^{2\nu-2}(t'')\,dt''\int_{t''}^\infty \frac{dt''}{a^{2\nu}(t'')}\nonumber\\&&
+p^2\int_{t_1}^\infty \frac{dt'_1}{a^{2\nu}(t'_1)}\int_{t_2}^\infty \frac{dt'_2}{a^{2\nu}(t'_2)}\int_{t'_1}^{t'_2}a^{2\nu-2}(t'')\,dt''+\dots\Bigg]\;.\nonumber\\&&{}
\end{eqnarray}
We see that the commutator of two fields vanishes essentially as $a^{-2\nu}$ for late times, and the same is true for the commutator of a field and its time derivative, but the commutators of two time derivatives arise only from the third terms in the expansions (31)--(33), and therefore go as $a^{-2\nu-2}$.
That is,
\begin{eqnarray*} 
\Big[\dot{\zeta}({\bf x}_1,t_1), \dot{\zeta}({\bf x}_2,t_2)\Big]&\rightarrow& 2i\,\int d^{2\nu}p\; {\rm Im}\Big[ C_p\zeta_p^{o*} \Big]\;e^{i{\bf p}\cdot({\bf x}_1-{\bf x}_2)}\\&&\times\Bigg[
\frac{p^2}{a^{2\nu}(t_1)\,\epsilon(t_1) a^{2\nu}(t_2)\,\epsilon(t_2)}\int_{t_1}^{t_2}a^{2\nu-2}(t')\epsilon(t')\,dt'+\dots\Bigg]\;,
\end{eqnarray*}
and likewise for $\gamma_{ij}$ and $\sigma_n$.

Let's now add up the total number of factors of $a(t_r),\; a(t_{r+1}),\;\dots {\rm and}\;a(t_N)$ in the integrand of Eq.~(2), for some selection of terms in the interactions  $H(t_s)$ with $r\leq s\leq N$.  Suppose that the selected term in $H(t_s)$  contains an explicit factor $a(t_s)^{A_s}$, and $B_s$ factors of field time derivatives.    Suppose also that in the inner $N-r+1$ commutators in Eq.~(2) there appear $C$ commutators of fields with each other, $C'$ commutators of fields with field time derivatives, and $C''$ commutators of field time derivatives with each other.
The number of field time derivatives that are {\em not} in these commutators is $\sum_s B_s-C'-2C''$, and these contribute a total $-2\sum_sB_s +2C'+4C''$ factors of $a$.  (All sums over $s$ here run from $r$ to $N$.)  In addition, there are $\sum_sA_s$ factors of $a$ that appear explicitly in the interactions, and as we have seen, the  commutators contribute $-2\nu C-2\nu C'-(2\nu+2)C''$ factors of $a$.  Hence the total number of factors of $a(t_r),\; a(t_{r+1}),\;\dots {\rm and}\;a(t_N)$ in the integrand of Eq.~(2) is 
\begin{equation}
\#=\sum_s(A_s-2B_s)-2\nu C-(2\nu-2)( C'+C'')=\sum_s(A_s-2B_s-2\nu+2)-2C\;,
\end{equation}
in which we have used the fact that the total number $C+C'+C''$ of commutators of the interactions $H(t_r),\; H(t_{r+1}),\;\dots {\rm and}\;H(t_N)$ with each other and with the field product $Q$ equals the number of these interactions.  Under the assumptions of this theorem,  all interactions have $A_s-2B_s\leq 2\nu-2$.  If any of them are safe in the sense that $A_s-2B_s<2\nu-2$, then $\#<0$, and the integral over time converges exponentially fast.  On the other hand, if all of them have $A_s-2B_s=2\nu-2$, then under the assumptions of this theorem they all involve only fields, not field time derivatives, so the same is true of the commutators of these interactions.  In this case $C>0$ and  $\#=-2C<0$, so again the integral over time converges exponentially fast.

In counting powers of $a$, we have held the wave numbers $p$ associated with internal lines fixed, like the external wave numbers, because we are integrating over time coordinates before we integrate over the internal wave numbers. The integrals over time receive little contribution from values of the conformal time $\tau\equiv -\int_t^\infty dt/a$ satisfying   $-p\tau\gg 1$ and $-q\tau\gg 1$, because of the rapid oscillation of the integrand, and for theories satisfying the conditions of our theorem they also receive little contribution from values of $\tau$ with $-p\tau\ll 1$ and $-q\tau\ll 1$, because of the damping provided by negative powers of $a$.   (Note that when $a(t)$ increases more or less exponentially, $\tau$ is of the order of $-1/aH$.)  Thus for these theories, we expect the integrals to be dominated by times for which $-1/\tau $ is  in the range from the $q$s to the $ps$s.  The question then is whether the integrals over the internal wave numbers $p$ are dominated by values of the order of the external wave numbers $q$?  If they are, then the results depend only on the history of inflation around the time of horizon exit, $-q\tau \approx 1$, or in other words, $q/a\approx H$.

Any integral over the internal wave numbers will in general take the form of a polynomial in the external wave numbers, with coefficients that may be divergent, plus a finite term given by a convergent integral dominated by internal wave numbers of the same order of magnitude as the fixed external wave numbers.  An example of this decomposition is given in Section VII.  In particular, the integral over the wave number associated with an internal line that begins and ends at the same vertex does not involve the external wave numbers, so its contribution is purely a polynomial in the wave numbers of the other lines attached to the same vertex. 

Just as in dealing with ultraviolet divergences in flat space quantum field theory, renormalization removes some of these ultraviolet divergent polynomial terms, and others are removed by appropriate redefinitions of the field operators.  (Some examples are given in the next section.)  Where redefinition of the field operators is necessary, it is only products of  the redefined ``renormalized'' field operators whose expectation values may be expected to give results that converge at late times.  If, after all such renormalizations and redefinitions, there remained ultraviolet divergences in the integrals over internal wave numbers, we could conclude that the approximation of extending the time integrals to $+\infty$ is not valid, and that these integrals can be taken only to some time $t$ late in inflation.  The decrease of the integrand at wave numbers $p$ much greater than $-1/\tau(t)$  would then provide the ultraviolet cut off that is still needed, but the correlation functions would  exhibit the sort of time dependence that has been found in other contexts by Woodard and his collaborators,$^3$ and we would not be able to draw conclusions about  correlations actually measured at times much closer to the present. The possible presence of such ultraviolet divergences that are not removed by renormalization and field redefinition is an important issue, which merits further study.\fnote{\dagger\dagger\dagger}{Many theories are afflicted with infrared divergences, even when $t$ is held fixed.  The infrared divergences are attributed to the imposition of the unrealistic initial condition, that at early times  all of infinite space is occupied by a Bunch--Davies vacuum.  The infrared divergence can be eliminated either by taking space to be finite$^8$ or by changing the vacuum.$^9$  In any case, it is the appearance of uncancelled ultraviolet rather than infrared divergences when we integrate over  internal wave numbers after taking the limit $t\rightarrow \infty$ that shows the impropriety of this interchange of limit and integral,  because factors of $1/a$ in the integrand are typically accompanied with factors of internal wave numbers, so that the $1/a$ factors do not suppress the integrand for large values of $a$ if the integral receives contributions from arbitrarily large values of the internal wave number.} But even if such ultraviolet divergences are present, it would still be possible to calculate the non-polynomial part of the integrals over internal momenta which is not ultraviolet divergent (at least in one loop order) even when the time $t$ is taken to infinity.  Such a calculation will be presented in Section VII.

\vspace{9pt}

\begin{center}
{\bf V. AN EXAMPLE: EXPONENTIAL EXPANSION}
\end{center}

To clarify the issues discussed at the end of the previous section, we will examine a simple  unphysical model, along with a revealing class of generalizations.  

First, consider a single real scalar field $\varphi({\bf x},t)$ in a fixed de Sitter metric.\fnote{\ddagger}{This model, and much of the analysis, was suggested to me by R. Woodard, private communication.}  In order to implement dimensional regularization, we work in $2\nu$ space dimensions, letting $\nu\rightarrow 3/2$ at the end of our calculation.  The Lagrangian density is taken as
\begin{equation} 
{\cal L}=-\frac{1}{2}\sqrt{-{\rm Det g}}\,g^{\mu\nu}\,(1+\lambda\varphi^2)\,\partial_\mu\varphi\,\partial_\nu\varphi=
(1+\lambda\varphi^2)\,\left[\frac{a^{2\nu}}{2}\dot{\varphi}^2-\frac{a^{2\nu-2}}{2}(\nabla\varphi)^2\right]\;,
\end{equation} 
where $a\propto e^{Ht}$ with $H$ constant.
(This of course can be rewritten as a free field theory, but it is instructive nonetheless, and will be generalized later in this section to interacting theories.)
We follow the usual procedure of defining a canonical conjugate field $\pi=\partial{\cal L}/\partial\dot{\varphi}$, constructing the Hamiltonian density ${\cal H}=\pi\dot{\varphi}-{\cal L}$ with $\dot{\varphi}$ expressed in terms of $\pi$, dividing ${\cal H}$ into a quadratic part ${\cal H}_0$ and interaction part ${\cal H}_I$, and then replacing $\pi$ in ${\cal H}_I$ with the interaction-picture $\pi_I$ given by $\dot{\varphi}=[\partial{\cal H}_0/\partial\pi]_{\pi=\pi_I}$.  This gives an interaction
\begin{equation} 
H_I=\frac{\lambda}{2} \int d^{2\nu}x\;\left[-\frac{a^2}{2}\left\{\frac{ \varphi^2}{1+\lambda\varphi^2}\,,\, \dot{\varphi}^2\right\}+a^{2\nu-2}(\nabla\varphi)^2\varphi^2\right]\;.
\end{equation} 
(An anticommutator is needed in the first term to satisfy the requirement that $H_I$ be Hermitian.)  This interaction satifies the conditions of the theorem proved in the previous section for any value of the space dimensionality $2\nu$: the first term in the square brackets contains $2\nu-4$ factors of $a$ (counting a factor $a^{-2}$ for each time derivative, so it is safe, while the second term contains $2\nu-2$ factors of $a$, and is therefore dangerous, but it only involves fields (including space derivatives), not their time derivatives, so though dangerous it still satisfies the conditions of our theorem.

To first order in $\lambda$, the expectation value $\langle \varphi({\bf x},t)\,\varphi({\bf x}',t)\rangle$ 
is given by a one-loop diagram, in which a scalar field line is emitted and absorbed at the same vertex, with the two external lines also attached to this vertex.  This expectation value receives contributions of three kinds:
\begin{description}
\item[i] Terms in which no time  derivatives act  on the internal lines.  This contribution is the same as would be obtained by adding  effective interactions proportional to $a^{2\nu-2}(\nabla \varphi)^2$,  $a^{2\nu-2}\varphi^2$, or $a^{2\nu}\dot{\varphi}^2$, all of which satisfy the conditions of the theorem of the previous section.  Thus it cannot affect the conclusion that the integral over the time argument of $H_I(t_1)$ converges exponentially at $t_1=+\infty$, so that   $\langle \varphi({\bf x},t)\,\varphi({\bf x}',t)\rangle$ approaches a finite limit for $t\rightarrow \infty$.
\item[ii] Terms in which time derivatives act on both ends of the internal line.  This produces an effective interaction proportional to $a^{2\nu}\varphi^2$, which violates the conditions of our theorem, but it can be removed by adding an $R\varphi^2\sqrt{-{\rm Det}g}$ counterterm in the Lagrangian.  (This cancellation is not automatic, because the condition of minimal coupling is not enforced by any symmetry.)
\item[iii] Terms in which a time derivative acts on just one end of the internal line.  This produces an effective interaction proportional to $a^{2\nu}\varphi\dot{\varphi}$, which violates the conditions of our theorem, and cannot be removed by adding a generally covariant counterterm to the Lagrangian.
\end{description}

To see in detail what trouble is caused by the third type of contribution, note that the interaction picture scalar field is given by a Fourier decomposition like Eq.~(16), with coefficient functions\fnote{\ddagger\ddagger}{Here and below we will not be careful to extend factors like $4\pi$ to $2\nu$ space dimensions.  This only affects constant terms that accompany any $(2\nu-3)^{-1}$ poles.}
\begin{equation} 
\varphi_q(t)=\frac{e^{i\pi(2\nu+1)/4}H^{\nu-1/2}}{4\pi\sqrt{2}q^\nu}H^{(1)}_\nu(-q\tau)\,(-q\tau)^\nu\;,
\end{equation}
where $\tau$ is the conformal time
\begin{equation}
\tau\equiv -\int^\infty_t \frac{dt'}{a(t')}=-\frac{1}{a(t)H}\;.
\end{equation}
The contribution of the third kind to the expectation value then has the Fourier transform
\begin{eqnarray}
&&\int d^{2\nu}x\; e^{i{\bf q}\cdot({\bf x}-{\bf x}')}\langle\varphi({\bf x},t)\,\varphi({\bf x}',t)\rangle_{\bf iii}=\left(\frac{H^{2\nu-1}}{32\pi^2}\right)^3\left(\frac{2\pi}{q}\right)^{4\nu}\nonumber\\&&
\times 4\pi\int_0^\infty \frac{dp}{p}\int_{-\infty}^t dt_1 \,a^{2\nu}(t_1)\left(\frac{d}{dt_1}\left|(-p\tau_1)^\nu H_\nu^{(1)}(-p\tau_1)\right|^2\right)\nonumber\\&&\times
{\rm Im}\frac{d}{dt_1}\left[\left((-q\tau_1)^\nu H_\nu^{(1)}(-q\tau_1)\right)^2
\left((-q\tau)^\nu H_\nu^{(1)}(-q\tau)\right)^{*2}\right]
\end{eqnarray}
 Let's see what happens if we evaluate this by integrating first over $p$ and then over $t_1$ from $-\infty$ to late times, or vice versa.  

To integrate first over $p$, we can change the variable of integration from $p$ to $z\equiv-p\tau_1$, in which case the first derivative with respect to $t_1$ can be replaced with
$d/dt_1=(z/a_1\tau_1)(d/dz)=-Hz(d/dz)$, while $dp/p=dz/z$.  Dimensional regularization (with $2\nu< 1$) makes the  function   $\left|z^\nu H_\nu^{(1)}(z)\right|$ vanish at $z\rightarrow \infty$, while for $\nu>0$ it takes the value $2^\nu\Gamma(\nu)/\pi$ for  $z\rightarrow 0$, so 
$$ \int_0^\infty dz\,\frac{d}{dz}\left|z^\nu H_\nu^{(1)}(z)\right|^2=-\left(\frac{2^\nu\Gamma(\nu)}{\pi}\right)^2\;,$$
and therefore 
\begin{eqnarray}
&&\int d^{2\nu}x\; e^{i{\bf q}\cdot({\bf x}-{\bf x}')}\langle\varphi({\bf x},t)\,\varphi({\bf x}',t)\rangle_{\bf iii}=-4\pi H\left(\frac{2^\nu\Gamma(\nu)}{\pi}\right)^2\left(\frac{H^{2\nu-1}}{32\pi^2}\right)^3\left(\frac{2\pi}{q}\right)^{4\nu}\nonumber\\&&
\times \int_{-\infty}^t dt_1 \,a^{2\nu}(t_1)
{\rm Im}\frac{d}{dt_1}\left[\left((-q\tau_1)^\nu H_\nu^{(1)}(-q\tau_1)\right)^2
\left((-q\tau)^\nu H_\nu^{(1)}(-q\tau)\right)^{*2}\right]\nonumber\\&&
\end{eqnarray}
For $t_1\rightarrow +\infty$ and $t\rightarrow +\infty$ (that is, $\tau\rightarrow 0$ and $\tau_1\rightarrow 0$), the integrand of the integral over $t_1$ on the second line has the constant limit 
\begin{eqnarray}&& a^{2\nu}(t_1)
{\rm Im}\frac{d}{dt_1}\left[\left((-q\tau_1)^\nu H_\nu^{(1)}(-q\tau_1)\right)^2
\left((-q\tau)^\nu H_\nu^{(1)}(-q\tau)\right)^{*2}\right]\rightarrow 
-\frac{4\Gamma(\nu)^2q^{2\nu}}{\pi^3H^{2\nu-1}}\;.\nonumber\\&&{}
\end{eqnarray} 
Thus for $t\rightarrow \infty$, the correlation function (39) does not approach a constant, 
but instead goes as \begin{eqnarray}
&&\int d^{2\nu}x\; e^{i{\bf q}\cdot({\bf x}-{\bf x}')}\langle\varphi({\bf x},t)\,\varphi({\bf x}',t)\rangle_{\bf iii}\rightarrow \frac{H^{4\nu-1}\Gamma(\nu)^4t}{2(2\pi)^{10-4\nu}q^{2\nu}}\;.
\end{eqnarray}
There is no pole here that prevents continuation to space dimensionality $2\nu=3$.  From this point of view, integrating first over $p$, the failure of the correlation function to approach a finite limit at late times is due to the fact already noted, that the integral over $p$ produces an effective interaction that does not satisfy the conditions of our theorem.

But suppose we first integrate over $t_1$ from $-\infty$ to $+\infty$.  Now there is no problem with convergence at late times, because the original interaction does satisfy the conditions of our theorem, but instead we now have a problem with the convergence of the integral over $p$.  It will be helpful to divide the integral over $p$ into an integral from $0$ to $\Lambda q$, where $\Lambda\gg 1$, and an integral from $\Lambda q$ to infinity.  The first integral obviously has no ultraviolet divergence, and the vanishing of the first time derivative in Eq.~(39) for $p\rightarrow 0$ prevents any infrared divergence.  In the second integral $p$ and  $-1/\tau$ are the only magnitudes in the problem with which $q$ can be compared, so for $t\rightarrow +\infty$ and hence $\tau\rightarrow 0$  we can evaluate the correlation function by letting $q\rightarrow 0$ and keeping only the leading term in $q$.  Here again we can use the limiting formula (41), now for $q\rightarrow 0$ instead of $\tau\rightarrow 0$ and $\tau_1\rightarrow 0$.  The integral over $t_1$ is then trivial, and we find that for $q\ll 1/\tau$
the correlation function is\begin{equation}
\int d^{2\nu}x\; e^{i{\bf q}\cdot({\bf x}-{\bf x}')}\langle\varphi({\bf x},t)\,\varphi({\bf x}',t)\rangle_{\bf iii}\rightarrow \frac{H^{4\nu-2}\Gamma(\nu)^4}{2(2\pi)^{10-2\nu}q^{2\nu}}\int_{\Lambda q}^\infty \frac{dp}{p}+{\rm finite}\;.
\end{equation}
The ultraviolet divergent integral over $p$ is the price we pay for the naughtiness of taking the limit $t\rightarrow \infty$ before we integrate over $p$.

In this model it is clear how to remedy the difficulty of calculating correlation functions at late times.  As already mentioned, the original Lagrangian density (35) actually describes a free field theory.  This is made manifest by defining a new scalar field
\begin{equation} 
\tilde{\varphi}\equiv \int \sqrt{1+\lambda\varphi^2}\, d\varphi \;,
\end{equation} 
for which the Lagrangian density takes the form
\begin{equation} 
{\cal L}=-\frac{1}{2}\sqrt{-{\rm Det g}}\,g^{\mu\nu}\,\partial_\mu\tilde{\varphi}\,\partial_\nu\tilde{\varphi}\;.
\end{equation} 
There is no problem in taking the late-time limit of the correlation function
$\int d^{2\nu}\,e^{i{\bf q}\cdot({\bf x}-{\bf x}')}\langle\tilde{\varphi}({\bf x},t)\,\tilde{\varphi}({\bf x}',t)\rangle$ --- it is  just $2^{2\nu}H^{2\nu-1}\Gamma(\nu)^2/32\pi^4q^{2\nu}$.  From this point of view, the growth of the correlation function (42) at late times is a result of our perversity in calculating the correlation function of $\varphi$ instead of $\tilde{\varphi}$.

Can we find fields whose correlation functions have a constant limit at late times in theories that satisfy the conditions of our theorem but are not equivalent to free field theories?  The general answer is not known, but here is a class of interacting field theories for which such ``renormalized'' fields can be found.   This time we 
consider an arbitrary number of real scalar fields $\varphi_n({\bf x},t)$ in a fixed de Sitter metric. The Lagrangian density is taken to have the form of a non-linear $\sigma$-model:
\begin{equation} 
{\cal L}=-\frac{1}{2}\sum_{nm}\sqrt{-{\rm Det g}}\,g^{\mu\nu}\,\Big(\delta_{nm}+\lambda K_{nm}(\varphi)\Big)\,\partial_\mu\varphi_n\,\partial_\nu
\varphi_m\;,
\end{equation} 
where $K_{nm}(\varphi)$ is an arbitrary real symmetric matrix function of the $\varphi_n$; $\lambda$ is a coupling constant; and again $a\propto e^{Ht}$ with $H$ constant.  The Hamiltonian derived from this Lagrangian density does satisfy the conditions of the theorem of Section IV, whatever the function $K_{nm}(\varphi)$.  

To first order in $\lambda$, the same problem discussed earlier in this section arises from  graphs in which an internal line of the  field $\varphi_n$ is emitted and absorbed from the same vertex, with a time derivative acting on just one end of this line.  Depending on what correlation function is being calculated, the contribution of such graphs is proportional to various contractions of partial derivatives of the function 
\begin{equation} 
A_{m}(\varphi)\equiv \sum_n \frac{\partial K_{nm}(\varphi)}{\partial \varphi_n}\;.
\end{equation} 
Suppose we make a redefinition of  the fields of first order in $\lambda$:
\begin{equation} 
\tilde{\varphi}_n\equiv \varphi_n-\lambda \Delta_n(\varphi)\;.
\end{equation} 
This changes the matrix $K$ to
\begin{equation} 
\tilde{K}_{nm}(\varphi)=K_{nm}(\varphi)+\frac{\partial \Delta_n(\varphi)}{\partial\varphi_m}+\frac{\partial \Delta_m(\varphi)}{\partial\varphi_n}\;,
\end{equation} 
and so 
\begin{equation} 
\tilde{A}_{m}(\varphi)\equiv \sum_n \frac{\partial \tilde{K}_{nm}(\varphi)}{\partial \varphi_n} =A_{m}(\varphi)+\sum_n\frac{\partial^2 \Delta_n(\varphi)}{\partial\varphi_n\partial\varphi_m}+\sum_n\frac{\partial \Delta_m(\varphi)}{\partial\varphi_n\partial\varphi_n}\;.
\end{equation}
Thus the fields $\tilde{\varphi}_n$ are renormalized, in the sense that to first order in $\lambda$ correlation functions have finite limits at late times, provided that 
\begin{equation}
\sum_n\frac{\partial^2 \Delta_n(\varphi)}{\partial\varphi_n\partial\varphi_m}+\sum_n\frac{\partial \Delta_m(\varphi)}{\partial\varphi_n\partial\varphi_n}=-A_{m}(\varphi)\;.
\end{equation}
This can be solved by first solving the Poisson equation
\begin{equation}
\sum_n\frac{\partial^2 B(\varphi)}{\partial\varphi_n\partial\varphi_n}
=-\frac{1}{2}\sum_n\frac{\partial A_n(\varphi)}{\partial\varphi_n}
\end{equation}
and then solving a second Poisson equation
\begin{equation}
\sum_n\frac{\partial^2 \Delta_m(\varphi)}{\partial\varphi_n\partial\varphi_n}
=-A_m(\varphi)-\frac{\partial B(\varphi)}{\partial\varphi_m}\;.
\end{equation}
Thus for at least to first order in this class of theories, it is always possible to find a suitable set of renormalized fields.  

Because  we can take the limit $t\rightarrow \infty$ only for the correlation functions of suitably defined fields (such as $\tilde{\varphi}_n$ in our example), the question  naturally arises, whether these are the fields whose correlation functions we want to calculate.  The answer is conditioned by the fact that astronomical observations of the cosmic microwave background or large scale structure are made following a long era that has intervened since the end of inflation, during which things happened about which we know  almost nothing, such as reheating, baryon and lepton synthesis, and dark matter decoupling.  The only thing that allows us to use observations to learn about inflation is that some quantities were conserved during this era, while fluctuation wave lengths were outside the horizon.  These are the only quantities whose correlation functions at the end of inflation can be interpreted in terms of current observations.  In the classical limit, the quantities that are conserved outside the horizon are $\zeta$ and $\gamma_{ij}$, but we don't know whether this will be true when quantum effects are taken into account.  Still, we can expect that quantities are conserved only when there is some symmetry principle that makes them conserved, and whatever symmetry principle keeps some quantity conserved from the end of inflation to the time of horizon re-entry is likely also to keep it conserved from the time of horizon exit to the end of inflation.  So we may guess that the quantities whose correlation functions we will need to know are just those whose correlation functions approach constant limits at the end of inflation.

\vspace{9pt}

\begin{center}
{\bf VI. DANGEROUS INTERACTIONS IN INFLATIONARY THEORIES}
\end{center}

We now return to the semi-realistic theories described in Section III.  We will show in this section that all interactions are of the type called for in the theorem of Section IV;  that is, they are all safe interactions that (in three space dimensions)  do not grow exponentially at late times (and in fact are suppressed at late times at least by a factor $a^{-1}$), or dangerous interactions containing only fields and not their time derivatives, which grow no faster that $a$ at late times.  Fortunately, as noticed by Maldacena$^2$ in a different context, for this purpose it is not necessary to solve the constraint equations (7) and (8), which are quite complicated especially when the $\sigma_n$ fields are included.  Inspection of these equations shows that when we count $\dot{\zeta}$, $\dot{\gamma}_{ij}$, and $\dot{\sigma}_n$ as of order $a^{-2}$, the 
auxiliary fields $N-1$ and $N^i$ are both also of order $a^{-2}$.\fnote{\ddagger\ddagger\ddagger}{In counting powers of $a$, note that the three-dimensional affine connection and Ricci tensor are independent of $a$, so the curvature scalar $R^{(3)}$ goes as $a^{-2}$.  For instance, for $\gamma_{ij}=0$, we have $R^{(3)}=-a^{-2}e^{-2\zeta}(4\nabla^2\zeta+2(\nabla \zeta)^2)$.}    This is apparent in the first-order solution (9) of the constraint equations, but it holds to all orders in the fields.  
To calculate the quantity $E^j{}_iE^i{}_j-(E^i{}_i)^2$ in Eq. (5), we note that
\begin{equation}
E^i{}_j=H\delta^i{}_j +\dot{\zeta}\delta^i{}_j+\frac{1}{2}\left[e^{-\gamma}\frac{\partial}{\partial t}e^\gamma\right]^i{}_j
-\frac{1}{2}\Big(\nabla^iN_j+\nabla_jN^i\Big)\;.
\end{equation}
The first term  $H\delta^i{}_j$ is of order zero in $a$, while all other terms are of order $a^{-2}$, so 
\begin{equation}
E^j{}_iE^i{}_j-(E^i{}_i)^2=-6H^2-12H\dot{\zeta}-4H\nabla_kN^k+O(a^{-4})
\end{equation}
(In deriving this result, we note that $\left[e^{-\gamma}\frac{\partial}{\partial t}e^\gamma\right]^i{}_i=\dot{\gamma}_{ii}=0$.)
The terms in (5) of first order in $N-1$ all cancel as a consequence of the constraint equation (8), while terms of second order in $N-1$ in Eq.~(5) (and in particular in $a^{3} e^{3\zeta}\dot{\bar{\varphi}}^2/2N$ and $-3H^3a^{3}e^{3\zeta}/2N$) are suppressed by at least a factor $a^{3}(a^{-2})^2$, and are therefore safe.  Therefore we can  isolate all  terms that are potentially dangerous by setting $N=1$, and find
\begin{eqnarray}
&&{\cal L}=\frac{a^{3}}{2}e^{3\zeta}\Bigg[R^{(3)}-2V(\bar{\varphi})
-6H^2-12H\dot{\zeta}-4H\nabla_kN^k\nonumber\\&&~~~~~~+\dot{\bar{\varphi}}^2-a^{-2}e^{-2\zeta}[\exp{(-\gamma)}]^{ij}\sum_n\partial_i\sigma_n\partial_j\sigma_n\Bigg]+O(a^{-1})\;,\nonumber\\&&{}
\end{eqnarray}
We note that $e^{3\zeta}\nabla_kN^k=\partial_k(e^{3\zeta}N^k)$, so this term vanishes when integrated over three-space, and therefore makes no contribution to the action.
The term proportional to $\dot{\zeta}$ can be written
$$ -6a^{3}e^{3\zeta}H\dot{\zeta}=\frac{\partial}{\partial t}\left(-2a^{3}\,H\,e^{3\zeta}\right)+a^{3}e^{3\zeta}\Big(6H^2+2\dot{H}\Big)\;.$$
The first term vanishes when integrated over time, so it gives no contribution to the action.  To evaluate the remaining terms we use the unperturbed inflaton field equation, which (with $8\pi G\equiv 1$) gives $\dot{H}=-\dot{\bar{\varphi}}^2/2$, and the Friedmann equation, which gives $6H^2=2V+\dot{\bar{\varphi}}^2$.  We then find a cancellation
$$ -V-3H^2+\frac{1}{2}\dot{\bar{\varphi}}^2+6H^2+2\dot{H}=0\;.$$
Aside from terms that make no contribution to the action, the Lagrangian density is then
\begin{equation}
{\cal L}=\frac{a^{3}}{2}e^{3\zeta}\Bigg[R^{(3)}-a^{-2}e^{-2\zeta}[\exp{(-\gamma)}]^{ij}\sum_n\partial_i\sigma_n\partial_j\sigma_n\Bigg]+O(a^{-1})\;.
\end{equation}
We see that, at least in this class of theories, the dangerous terms that are not suppressed by a factor $a^{-1}$ grow at most like $a$ at late times, and involve only fields, not their time derivatives, as assumed in the theorem of Section III.  
	
It remains to be seen if in these theories, after integrating over times and taking the limit $t\rightarrow\infty$,
the remaining integrals over internal wave numbers are made convergent by the same counterterms that eliminate ultraviolet divergences in flat spacetime, and if not, whether they can be made convergent by suitable redefinitions of the fields $\zeta$ and $\gamma_{ij}$ appearing in the correlation functions.  This is left as a problem for further work.

Not all theories satisfy the conditions of the theorem of Section IV.  For instance, a  non-derivative interaction $\sqrt{-{\rm Det}g}F(\sigma)$ of the $\sigma $ fields would have $+3$ factors of $a$,  and hence would violate the condition that the total number of factors of $a$ (counting each time derivative as -2 factors) must be no greater than $+1$.  The $\sigma$ fields must be the Goldstone bosons of some broken global symmetry in order to satisfy the conditions of our theorem in a natural way.

\begin{center}
{\bf VII. A SAMPLE CALCULATION}
\end{center}

As an application of the formalism described in this paper, we will now calculate the one-loop contribution to the correlation function of two $\zeta$ fields, which is measured in the spectrum of anisotropies of the cosmic microwave background.  As already mentioned, in the class of theories described in Section III, this two-point function is dominated by a matter loop, because there are many types of matter field and only one
gravitational field.  We first consider the contribution of second order in the interaction (27).  It saves a great deal of work if we use the interaction-picture field equations (11) and (13) to put this interaction  in the form
\begin{equation} 
H_{\zeta\sigma\sigma}(t)=-\int d^{3} x\;{\cal L}_{\zeta\sigma\sigma}({\bf x},t)=A(t)+\dot{B}(t)
\end{equation} 
where
\begin{eqnarray} 
A&=&-2\epsilon H a^5 \sum_n\int d^{3} x\;\dot{\sigma}_n^2\,\nabla^{-2}\dot{\zeta}\\
B&=& \sum_n\int d^{3} x\;\left(\frac{a\zeta}{H}-\epsilon a^{3}\nabla^{-2}\dot{\zeta}\right)\,\left(\frac{1}{2}(\nabla \sigma_n)^2+\frac{1}{2}a^2\dot{\sigma}_n^2\right)\;.
\end{eqnarray} 
In general, for an interaction Hamiltonian of the form (58), Eq. (2) can be put in the form
\begin{eqnarray}
\langle Q(t)\rangle &=&\sum_{N=0}^\infty i^N\, \int_{-\infty}^t dt_N \int_{-\infty}^{t_N} dt_{N-1} \cdots \int_{-\infty}^{t_2} dt_1 \nonumber\\&&\times\left\langle \Big[\tilde{H}_I(t_1),\Big[\tilde{H}_I(t_2),\cdots \Big[\tilde{H}_I(t_N),\tilde{Q}^I(t)\Big]\cdots\Big]\Big]\right\rangle\;,
\end{eqnarray}
where
\begin{equation} 
\tilde{H}_I(t)=e^{iB(t)}\Big[A(t)+\dot{B}(t)+i e^{-iB(t)}\left(\frac{d}{dt} e^{iB(t)}\right)\Big] e^{-iB(t)}=A(t)+i[B(t),A(t)]+\frac{i}{2}[B(t),\dot{B}(t)]+\dots
\end{equation} 
\begin{equation} 
\tilde{Q}^I(t)= e^{iB(t)}Q^I(t) e^{-iB(t)}=Q^I(t)+i[B(t),Q^I(t)]-\frac{1}{2}[B(t),[B(t),Q^I(t)]]+\dots\;.
\end{equation} 
To second order in an  interaction of the form (58), the expectation value is then
\begin{eqnarray}
\langle Q(t)\rangle_2 &=&- \int_{-\infty}^t dt_2 \int_{-\infty}^{t_2} dt_1 \left\langle \Big[A(t_1),\Big[A(t_2),{Q}^I(t)\Big]\Big]\right\rangle
\nonumber\\&&-  \int_{-\infty}^t dt_1\;\left\langle\Big[\Big[B(t_1), A(t_1)+\dot{B}(t_1)/2\Big],Q^I(t)\Big]\right\rangle
\nonumber\\&&-\left\langle\Big[B(t),\Big[B(t),Q^I(t)\Big]\Big]\right\rangle 
\;,
\end{eqnarray}
The Fourier transform of the second-order term in the expectation value of a  product of two $\zeta$s is then
\begin{eqnarray}
&&\int d^{3}x \,e^{i{\bf q}\cdot( {\bf x}-{\bf x}')}\Big\langle {\rm vac,in}\Big|
\zeta ({\bf x},t)\,\zeta ({\bf x}',t)\Big|{\rm vac,in}\Big\rangle_2\nonumber\\&&
=-\frac{32(2\pi)^9}{q^4}{\rm Re}\int_{-\infty}^t a^5(t_2)\,\epsilon(t_2)\,H(t_2)\,dt_2
\nonumber\\&&~~~\times\;\int_{-\infty}^{t_2} a^5(t_1)\,\epsilon(t_1)\,H(t_1)\,dt_1\nonumber\\&&
\times\,\dot\zeta _q(t_1)\,\zeta ^*_q(t)\,\Big(\dot\zeta_q(t_2)\,\zeta^*_q(t)\,-\zeta_q(t)\,\dot\zeta^*_q(t_2)\,\Big)\nonumber\\&&
\times\,{\cal N}\,\int d^{3}p\int d^{3}p'\; \delta^{3}({\bf p}+{\bf p}'+{\bf q})\nonumber\\&&
~~~\times \;\dot{\sigma}_{p}(t_1)\,\dot{\sigma}^*_{p}(t_2)\,\dot{\sigma}_{p'}(t_1)\,\dot{\sigma}^*_{p'}(t_2)
\nonumber\\&&
+\frac{(2\pi)^{3}}{4q^4}{\cal N}\,\int d^{3}p\int d^{3}p'\; \delta^{3}({\bf p}+{\bf p}'+{\bf q})\nonumber\\&&
\times\,({\bf p}\cdot {\bf p}')^2\,|\sigma_{p}(t)|^2\,|\sigma_{p'}(t)|^2
\nonumber\\&&
+ \;\dots
\end{eqnarray}
where ${\cal N}$ is the number of $\sigma$ fields.  We have shown here explicitly the contribution of the first and third lines on the right-hand side of Eq.~(64).  The dots represent  one-loop contributions of the second line,  in which $[B,A+\dot{B}/2]$ plays the role of a $\zeta\zeta\sigma\sigma$ ``seagull'' interaction, as well  as one-loop terms of first order in the $\zeta\zeta\sigma\sigma$ terms in Eq.~(5), in both of which the integral over internal wave number is ${\bf q}$-independent, plus counterterms arising in first order from interactions that cancel ultraviolet divergences in flat space, including $\sqrt{-{\rm Det} g}\,R^{\mu\nu}R_{\mu\nu}$ and $\sqrt{-{\rm Det} g}\,R^2$ terms in the Lagrangian density that are not included in Eq.~(5). 

Though it has not been made explicit in this section, we use dimensional regularization to remove infinities in the integrals over $p$ and $p'$ at intermediate stages in the calculation, and we now assume that the singularity as the number of space dimensions approaches three is cancelled by the terms in Eq.~(65) represented by dots, leaving it to future work to show that this is the case.   Then these integrals are dominated by $p\approx p'\approx q$.  As we have seen, the integrals over time are then dominated by the  time $t_q$ of horizon exit, when $q/a(t_q)\simeq H(t_q)$.  For simplicity, we will assume (for the first time in this paper) that the unperturbed inflaton field $\bar{\varphi}(t)$ is rolling very slowly down the potential at time $t_q$, so that the expansion near this time can be approximated as strictly exponential, $a(t)\propto e^{Ht}$.  Then the wave functions are
$$\sigma_q(t)\simeq \sigma_q^o\,e^{-iq\tau}\Big(1+iq\tau\Big)\;,$$
$$\zeta_q(t)\simeq \zeta^o_q \,e^{-iq\tau}\Big(1+iq\tau\Big)\;,$$
where $\tau$ is the conformal time
$$\tau\equiv -\int_t^\infty \frac{dt}{a(t)}\;,$$
and the wave functions outside the horizon have modulus
$$|\sigma_q^o|^2=\frac{H^2(t_q)}{2(2\pi)^{3}\,q^{3}}\;,~~~|\zeta_q^o|^2=\frac{H^2(t_q)}{2(2\pi)^{3}\,\epsilon(t_q)\,q^{3}}$$
Using these wave functions in Eq.~(65) gives
\begin{eqnarray}
&&\int d^{3}x \,e^{i{\bf q}\cdot( {\bf x}-{\bf x}')}\Big\langle {\rm vac,in}\Big|
\zeta({\bf x},t)\,\zeta({\bf x}',t)\Big|{\rm vac,in}\Big\rangle_2\nonumber\\&&
=\frac{(8\pi G H^2(t_q))^2{\cal N}}{(2\pi)^{3}}\int d^{3}p \int d^{3}p'\;\delta^{3}({\bf p}+{\bf p}'+{\bf q})\nonumber\\&&~~~
\times\,\left[\frac{p\,p'}{q^7\,(p+p'+q)}+\frac{({\bf p}\cdot{\bf p}')^2}{16\,q^4\,p^{3}\,p'^{3}}\right]+\dots
\end{eqnarray}
with the dots having the same meaning as in Eq.~(65).  

Simple dimensional analysis tells us that when the integral over internal wave numbers of the first term in square brackets is made finite by dimensional regularization, it is converted to
\begin{equation} 
 \int d^{3}p \int d^{3}p'\;\delta^{3}({\bf p}+{\bf p}'+{\bf q})\frac{p\,p'}{p+p'+q}
\Rightarrow q^{4+\delta}F(\delta)\;,
\end{equation}  
where $\delta$ is a measure of the difference between the space dimensionality and three.  The ultraviolet divergences in this integrals for $\delta=0$ gives the function $F(\delta)$ a singularities as $\delta\rightarrow 0$:
\begin{equation}
F(\delta)\rightarrow \frac{F_0}{\delta}+F_1\;,
\end{equation} 
so that in the limit $\delta=0$
\begin{equation} 
 \int d^{3}p \int d^{3}p'\;\delta^{3}({\bf p}+{\bf p}'+{\bf q})\frac{p\,p'}{p+p'+q}
= q^4 \,\Big[F_0\ln q+L\Big]\;,
\end{equation}
where $L$ is a  divergent constant.  
We can easily calculate the coefficient $F_0$ of the logarithm. For this purpose, we note that, in general,
\begin{equation}
\int d^{3}p\int d^{3}p'\; \delta^{3}({\bf p}+{\bf p}'+{\bf q})f(p,p',q)
=\frac{2\pi}{q}\int_0^\infty p\,dp\int_{|p-q|}^{p+q} p'\,dp'\;f(p,p',q)
\end{equation}
To eliminate the divergence in the integral over $p$ and $p'$, we multiply by $q$ and differentiate
six times with respect to $q$.  A tedious but straightforward calculation gives
$$
\frac{d^6}{dq^6}\left[q\int d^{3}p \int d^{3}p'\;\delta^{3}({\bf p}+{\bf p}'+{\bf q})\frac{p\,p'}{p+p'+q}\right]=-\frac{8\pi}{q}
$$
Comparing this with the result of applying the same operation to Eq.~(69) then gives $F_0=-\pi/15$.

In contrast, the integral of the second term in square brackets in Eq.~(66) is a sum of powers of $q$ with divergent coefficients, but with no logarithmic singularity in $q$.  (This term would be eliminated if we calculated the expectation value of a product of fields $\tilde{\zeta}\equiv \exp(-iB)\zeta\exp(iB)$ instead of $\zeta$.) The terms represented by dots in Eq.~(65) 
make contributions that are also just a sum of powers of $q$ with divergent coefficients.  We are assuming that all ultraviolet divergences cancel, but we cannot find resulting finite power terms without knowing the renormalized coefficients of the $\sqrt{-{\rm Det} g}R^{\mu\nu}R_{\mu\nu}$ and $\sqrt{-{\rm Det} g}R^2$ terms in the Lagrangian density.  So we are left with the result (now restoring a suitable power of $8\pi G$) that
\begin{eqnarray}
&&\int d^{3}x \,e^{i{\bf q}\cdot( {\bf x}-{\bf x}')}\Big\langle {\rm vac,in}\Big|
\zeta({\bf x},t)\,\zeta({\bf x}',t)\Big|{\rm vac,in}\Big\rangle_2\nonumber\\&&
=-\frac{\pi\Big(8\pi G H^2(t_q)\Big)^2{\cal N}}{15  (2\pi)^{3}q^{3}}\Big[\ln q+C \Big]
\end{eqnarray}
with $C$ an unknown constant. This may be compared with the classical (and classic) result, that in slow roll inflation this correlation function takes the form
\begin{eqnarray}
&&\int d^{3}x \,e^{i{\bf q}\cdot( {\bf x}-{\bf x}')}\Big\langle {\rm vac,in}\Big|
\zeta({\bf x},t)\,\zeta({\bf x}',t)\Big|{\rm vac,in}\Big\rangle_0\nonumber\\&&
=\frac{8\pi G H^2(t_q)}{4 (2\pi)^3|\epsilon(t_q)|q^3}
\end{eqnarray}
The one-loop correction (71) is smaller by a factor of order $8\pi G H^2{\cal N}|\epsilon(t_q)|$, so even if ${\cal N}$ is $10^2$ or $10^3$ this correction is likely to remain unobservable.  Still, it is interesting that even in the extreme slow roll limit, where $H(t_q)$ and $\epsilon(t_q)$ are nearly constant, the factor $\ln q$ gives it a different dependence on the wave number $q$.

\vspace{12pt}

\begin{center}
{\bf ACKNOWLEDGMENTS}
\end{center}

For helpful conversations I am grateful to K. Chaicherdsakul, S. Deser, W. Fischler, E. Komatsu, J. Maldacena, A. Vilenkin, and R. Woodard.  This material is based upon work supported by the National Science Foundation under Grants Nos. PHY-0071512 and PHY-0455649 and with support from The Robert A. Welch Foundation, Grant No. F-0014, and also grant support from the US Navy, Office of Naval Research, Grant Nos. N00014-03-1-0639 and N00014-04-1-0336, Quantum Optics Initiative.

\pagebreak

\renewcommand{\theequation}{A.\arabic{equation}}
\setcounter{equation}{0}

\begin{center}
{\bf APPENDIX: THE IN-IN FORMALISM}
\end{center}

\noindent
{\bf 1. Time Dependence}

First, it is necessary to be precise about the origin of the time-dependence of the fluctuation Hamiltonian in applications such as those encountered in cosmology.   Consider a general Hamiltonian system, with canonical variables $\phi_a({\bf x},t)$ and conjugates $\pi_a({\bf x},t)$ satisfying the commutation relations
\begin{equation} 
\Big[\phi_a({\bf x},t),\pi_b({\bf y},t)\Big]=i\delta_{ab}\delta^{3}({\bf x}-{\bf y})\;,~~~~~\Big[\phi_a({\bf x},t),\phi_b({\bf y},t)\Big]= \Big[\pi_a({\bf x},t),\pi_b({\bf y},t)\Big]=0\;,
\end{equation} 
and the equations of motion
\begin{equation} 
\dot{\phi}_a({\bf x},t)=i\Big[H[\phi(t),\pi(t)],\phi_a({\bf x},t)\Big] \;,~~~~~~~ \dot{\pi}_a({\bf x},t)=i\Big[H[\phi(t),\pi(t)],\pi_a({\bf x},t)\Big] \;.
\end{equation} 
Here $a$ is a compound index labeling particular fields and their spin components.  
The Hamiltonian $H$ is a functional of the $\phi_a({\bf x},t)$ and  $\pi_a({\bf x},t)$ at fixed time $t$, which according to Eq.~(A.2) is of course
independent of the time at which these variables are evaluated.  

We assume the existence of a time-dependent c-number solution $\bar{\phi}_a({\bf x},t)$, 
$\bar{\pi}_a({\bf x},t)$, satisfying the classical equations of motion:
\begin{equation} 
\dot{\bar{\phi}}_a({\bf x},t)=\frac{\delta H[\bar{\phi}(t),\bar{\pi}(t)]}{\delta \bar{\pi}_a({\bf x},t)}\;,~~~~~
\dot{\bar{\pi}}_a({\bf x},t)=-\frac{\delta H(\bar{\phi}(t),\bar{\pi}(t)]}{\delta \bar{\phi}_a({\bf x},t)}\;,
\end{equation} 
and we expand around this solution, writing
\begin{equation} 
\phi_a({\bf x},t)=\bar{\phi}_a({\bf x},t)+\delta\phi_a({\bf x},t)\;,~~~~~~~~~~
\pi_a({\bf x},t)=\bar{\pi}_a({\bf x},t)+\delta\pi_a({\bf x},t)\;.
\end{equation} 
(In cosmology, $\bar{\phi}_a$ would describe the Robertson--Walker metric and the expectation values of various scalar fields.)
Of course, since c-numbers commute with everything, the fluctuations satisfy the same commutation rules (A.1) as the total variables:
\begin{equation} 
\Big[\delta\phi_a({\bf x},t),\delta\pi_b({\bf y},t)\Big]=i\delta_{ab}\delta^{3}({\bf x}-{\bf y})\;,~~~~~\Big[\delta\phi_a({\bf x},t),\delta\phi_b({\bf x},t)\Big]= \Big[\delta\pi_a({\bf x},t),\delta\pi_b({\bf x},t)\Big]=0\;,
\end{equation}
When the Hamiltonian is expanded in powers of the perturbations $\delta\phi_a({\bf x},t)$ and $\delta\pi_a({\bf x},t)$ at some definite time $t$, we encounter terms of zeroth and first order in the perturbations, as well as time-dependent terms of second and higher order:
\begin{eqnarray} 
H[\phi(t),\pi(t)]&=&H[\bar{\phi}(t),\bar{\pi}(t)]+ \sum_a\frac{\delta H[\bar{\phi}(t),\bar{\pi}(t)]}{\delta \bar{\phi}_a({\bf x},t)}\delta\phi_a({\bf x},t]
+\sum_a\frac{\delta H[\bar{\phi}(t),\bar{\pi}(t)]}{\partial \bar{\pi}_a({\bf x},t)}\delta\pi_a({\bf x},t)
\nonumber\\&&~~~~~~+\tilde{H}[\delta\phi(t),\delta\pi(t);t]\;,
\end{eqnarray} 
where $\tilde{H}[\delta\phi(t),\delta\pi(t);t]$ is the sum of all terms in $H[\bar{\phi}(t)+\delta\phi(t), \bar{\pi}(t)+\delta\pi(t)]$ of second and higher order in the $\delta\phi({\bf x},t)$ and/or $\delta\pi({\bf x},t)$.  

Now, although $H$ generates the time-dependence of $\phi_a({\bf x},t)$ and $\pi_a({\bf x},t)$, it is $\tilde{H}$ rather than $H$ that generates the time dependence of $\delta\phi_a({\bf x},t)$ and $\delta\pi_a({\bf x},t)$.
That is, Eq.~(A.2) gives
$$
\dot{\bar{\phi}}_a({\bf x},t)+\delta\dot{\phi}_a({\bf x},t)=i\Big[H[\phi(t),\pi(t)],\delta\phi_a({\bf x},t)\Big] \;,~~~~~~ \dot{\bar{\pi}}_a({\bf x},t)+\delta\dot{\pi}_a({\bf x},t)=i\Big[H[\phi(t),\pi(t)],\delta\pi_a({\bf x},t)\Big] \;,
$$
while Eqs.~(A.5) and (A.3) give
$$
i\left[\sum_b\int d^{3}y\,\frac{\delta H[\bar{\phi}(t),\bar{\pi}(t)]}{\delta \bar{\phi}_b({\bf y},t)}\delta\phi_b({\bf y},t)
+\sum_b\int d^{3}y\,\frac{\delta H[\bar{\phi}(t),\bar{\pi}(t)]}{\delta \bar{\pi}_b({\bf y},t)}\delta\pi_b({\bf y},t),
\delta\phi_a({\bf x},t)\right]=\dot{\bar{\phi}}_a({\bf x},t)
$$
$$
i\left[\sum_b\int d^{3}y\,\frac{\delta H[\bar{\phi}(t),\bar{\pi}(t)]}{\delta \bar{\phi}_b({\bf y},t)}\delta\phi_b({\bf y},t)
+\sum_b\int d^{3}y\,\frac{\delta H[\bar{\phi}(t),\bar{\pi}(t)]}{\delta \bar{\pi}_b({\bf y},t)}\delta\pi_b({\bf y},t),
\delta\pi_a({\bf x},t)\right]=\dot{\bar{\pi}}_a({\bf x},t)\;.
$$
Subtracting, we find
\begin{equation} 
\delta\dot{\phi}_a({\bf x},t)=i\Big[\tilde{H}[\phi(t),\pi(t);t],\delta\phi_a({\bf x},t)\Big] \;,~~~~~~~ \delta\dot{\pi}_a({\bf x},t)=i\Big[\tilde{H}[\phi(t),\pi(t);t],\delta\pi_a({\bf x},t)\Big] \;.
\end{equation}
This then is our prescription for constructing the time-dependent Hamiltonian $\tilde{H}$ that governs the time-dependence of the fluctuations: expand the original Hamiltonian $H$ in powers of fluctuations $\delta\phi$ and $\delta\pi$, and throw away the terms of zeroth and first order in these fluctuations.  It is this construction that gives $\tilde{H}$ an explicit dependence on time.

\vspace{9pt}

\noindent
{\bf 2. Operator Formalism for Expectation Values}

We  consider a general Hamiltonian system, of the sort described in the previous subsection. 
It follows from Eq.~(A.7) that the fluctuations at time $t$ can be expressed in terms of the same operators at some very early time $t_0$ through a unitary transformation
\begin{equation}
\delta\phi_a(t)=U^{-1}(t,t_0)\delta\phi_a(t_0)\,U(t,t_0)\;,~~~~
\delta\pi_a(t)=U^{-1}(t,t_0)\delta\pi_a(t_0)\,U(t,t_0)\;,
\end{equation}
where $U(t,t_0)$ is defined by the differential equation
\begin{equation}
\frac{d}{dt}U(t,t_0)=-i\,\tilde{H}[\delta\phi(t),\delta\pi(t);t]\,U(t,t_0)
\end{equation}
and the initial condition
\begin{equation}
U(t_0,t_0)=1\;.
\end{equation}
In the application that concerns us in cosmology, we can take $t_0=-\infty$, by which we mean any time early enough so that the wavelengths of interest are deep inside the horizon.

To calculate $U(t,t_0)$, we now further decompose $\tilde{H}$ into a kinematic term $H_0$ that is quadratic in the fluctuations, and an interaction term $H_I$:
\begin{equation}
\tilde{H}[\delta\phi(t),\delta\pi(t);t]=H_0[\delta\phi(t),\delta\pi(t);t]+H_I[\delta\phi(t),
\delta\pi(t);t]\;,
\end{equation}
and we seek to calculate $U$ as a power series in $H_I$.  To this end, we introduce an ``interaction picture'': we define fluctuation operators $\delta\phi^I_a(t)$ and $\delta\pi^I_a(t)$ whose time 
dependence is generated by the quadratic part of the Hamiltonian:
\begin{equation} 
\delta\dot{\phi}^I_a(t)=i\Big[H_0[\delta\phi^I(t),\delta\pi^I(t);t],\delta\phi^I_a(t)\Big] \;,~~~~~~~ \delta\dot{\pi}^I_a(t)=i\Big[H_0[\delta\phi^I(t),\delta\pi^I(t);t],\delta\pi^I_a(t)\Big] \;,
\end{equation}
and the initial conditions
\begin{equation}
\delta{\phi}^I_a(t_0)=\delta{\phi}_a(t_0)\;,~~~~~
\delta{\pi}^I_a(t_0)=\delta{\pi}_a(t_0)\;.
\end{equation}
Because $H_0$ is quadratic, the interaction picture operators are free fields, satisfying linear wave equations.

It follows from Eq.~(A.12) that in evaluating $H_0[\delta\phi^I,\delta\pi^I;t]$ we can take the time argument of $\delta\phi^I$ and $\delta\pi^I$ to have any value, and in particular we can take it as $t_0$, so that
\begin{equation}
H_0[\delta\phi^I(t),\delta\pi^I(t);t]=H_0[\delta\phi(t_0),\delta\pi(t_0);t]\;,
\end{equation}
but the intrinsic time-dependence of $H_0$ still remains.  The solution of Eq.~(A.12) can again be written as a unitary transformation:
\begin{equation}
\delta\phi_a^I(t)=U^{-1}_0(t,t_0)\delta\phi_a(t_0)U_0(t,t_0)\;,~~~~
\delta\pi_a^I(t)=U^{-1}_0(t,t_0)\delta\pi_a(t_0)U_0(t,t_0)\;,
\end{equation}
with $U_0$ defined by the differential equation
\begin{equation}
\frac{d}{dt}U_0(t,t_0)=-i\,H_0[\delta\phi(t_0),\delta\pi(t_0);t]\,U_0(t,t_0)
\end{equation}
and the initial condition
\begin{equation}
U_0(t_0,t_0)=1\;.
\end{equation}
Then from Eqs.~(A.9) and (A.16) we have
$$
\frac{d}{dt}\Big[U_0^{-1}(t,t_0)U(t,t_0)\Big]=-iU_0^{-1}(t,t_0)H_I[\delta\phi(t_0),\delta\pi(t_0);t]U(t,t_0)\;.
$$
Using Eq.~(A.15), this gives
\begin{equation}
U(t,t_0)=U_0(t,t_0)F(t,t_0)\;,
\end{equation}
where
\begin{equation}
\frac{d}{dt}F(t,t_0)=-iH_I(t)F(t,t_0)\;,~~~~F(t_0,t_0)=1\;.
\end{equation}
and $H_I(t)$ is the interaction Hamiltonian in the interaction picture:
\begin{equation}
H_I(t)\equiv U_0(t,t_0)H_I[\delta\phi(t_0),\delta\pi(t_0);t]U_0^{-1}(t,t_0)=H_I[\delta\phi^I(t),\delta\pi^I(t);t]
\end{equation}
The solution of equations like (A.19) is well known
\begin{equation}
F(t,t_0)=T\exp\left(-i\int_{t_0}^t H_I(t)\,dt\right)
\end{equation}
where $T$ indicates that the products of $H_I$s in the power series expansion of the exponential are to be time-ordered; that is, they are to be written from left to right in the decreasing order of time arguments.  The solution for the fluctuations in terms of the free fields of the interaction picture is then given by Eqs.~(A.8) and (A.15) as
\begin{eqnarray}
 Q(t)  &=& F^{-1}(t,t_0)\,Q^I(t)F(t,t_0) \nonumber\\&=& \left[\bar{T}\exp\left(i\int_{t_0}^t H_I(t)\,dt\right)\right]\,Q^I(t)\,\left[T\exp\left(-i\int_{t_0}^t H_I(t)\,dt\right)\right]\;,
\end{eqnarray}
where $Q(t)$ is any $\delta\phi({\bf x},t)$ or $\delta\pi({\bf x},t)$ or any product of the $\delta\phi$s
and/or $\delta\pi$s, all at the same time $t$ but in general with different space coordinates, and $Q^I(t)$ is the same product of $\delta\phi^I({\bf x},t)$ and/or $\delta\pi^I({\bf x},t)$.  Also, $\bar{T}$ denotes anti-time-ordering: products of $H_I$s in the power series expansion of the exponential  are to be written from left to right in the {\em increasing} order of time arguments. 

\vspace{9pt}

\noindent
{\bf 3. Diagrammatic Formalism for Expectation Values}

We want to use Eq.~(A.22) to calculate the expectation value $\langle Q(t)\rangle $of the product $Q(t)$ in a ``Bunch--Davies'' vacuum, annihilated by the positive-frequency part of the interaction picture fluctuations $\delta\varphi^I$ and $\delta\pi^I$.  We can  use the familiar Wick theorem to express the vacuum expectation value of the right-hand side of Eq.~(A.22) as a sum over pairings of the $\delta\varphi^I$ and $\delta\pi^I$ with each other.  (This of course is the same as supposing the interaction-picture fields in $H_I(t)$ and $Q^I(t)$ to be governed by a Gaussian probability distribution, except that the order of operators in bilinear averages has to be the same as the order in which they appear in Eq.~(A.22).)   Expanding Eq.~(A.22) as a sum of products of bilinear products leads to a set of diagrammatic rules, but one that is rather complicated.  

In calculating the term in  $\langle Q \rangle$  of $N$th order in the interaction, we draw all  diagrams with $N$ vertices.  Just as for ordinary Feynman diagrams, each vertex is labeled with a space and time coordinate, and has lines attached corresponding to the fields in the interaction.
There are also external lines, one for each field operator in the product $Q$, labeled with the different space coordinates and the common time $t$ in the arguments of these fields.
All external lines are connected to vertices or other external lines, and all remaining lines attached to vertices are attached to other vertices.  But there are significant differences between the rules following from Eq.~(A.22) and the usual Feynman rules:
\begin{itemize}
\item   We have to distinguish between ``right'' and ``left'' vertices, arising respectively from the time-ordered product and the anti-time-ordered product.  A diagram with $N$ vertices contributes a sum over all $2^N$ ways of choosing each vertex to be a left vertex or a right vertex.  Each right or left vertex contributes a factor $-i$ or $+i$, respectively, as well as whatever coupling parameters appear in the interaction.
\item A line connecting two right vertices or a right vertex and an external line, in which it is associated with  field operators $A({\bf x},t')$ and $B({\bf y},t'')$, contributes a conventional Feynman propagator $\langle T\{A({\bf x},t')B({\bf y},t''\}\rangle$.
(It will be understood here and below, that in calculating propagators all fields $A$, $B$, etc. are taken in the interaction picture, and can be $\delta\varphi^I$s and/or $\delta\pi^I$s.)  As a special case, if $B$ is associated with an external line then $t''=t$, and since $t'\leq t$, this is $\langle B({\bf y},t)A({\bf x},t')\rangle$. 
\item A line connecting two left vertices, associated with  field operators $A({\bf x},t')$ and $B({\bf y},t'')$, contributes a  propagator $ \langle\bar{T}\{A({\bf x},t')B({\bf y},t''\}\rangle$.  As a special case, if $B$ is associated with an external line then $t''=t$, and this is $\langle A({\bf x},t')B({\bf y},t)\rangle$.  
\item A line connecting a left vertex, in which it is associated with a field operator $A({\bf x},t')$, to a right vertex, in which it is associated with a field operator $B({\bf y},t'')$, contributes a propagator $\langle A({\bf x},t')B({\bf y},t'')\rangle$.
\item We must integrate over all over the times $t',\,t'',\dots$, associated with the vertices from $t_0$ to $t$, as well as  over all space coordinates associated with the vertices. 
\end{itemize}

We must say a word about the disconnected parts of diagrams.  A vacuum fluctuation subdiagram is one in which each vertex is connected only to other vertices, not to external lines.  Just as in ordinary quantum field theories, the sum of all vacuum fluctuation diagrams  contributes a numerical factor multiplying the contribution of diagrams in which vacuum fluctuations are excluded.  But unlike the case of ordinary quantum field theory, this numerical factor is not a phase factor, but is simply
\begin{equation}
\left\langle \left[\bar{T}\exp\left(i\int_{t_0}^t H_I(t)\,dt\right)\right]\,\left[T\exp\left(-i\int_{t_0}^t H_I(t)\,dt\right)\right]\right\rangle=1\;.
\end{equation}
Hence in the ``in-in'' formalism all vacuum fluctuation diagrams automatically cancel.  Even so, a diagram may contain disconnected parts which do not cancel, such as external lines passing through the diagram without interacting.  Ignoring  all disconnected parts gives what in the theory of noise is known as the {\em cumulants} of expectation values,$^{10}$ from which the full expectation values can easily be calculated as a sum of products of cumulants.

\vspace{9pt}

\noindent
{\bf 4. Path Integral Derivation of the Diagrammatic Rules}.

It is often preferable use path integration instead of the operator formalism,  in order to derive the Feynman rules directly from the Lagrangian rather than from the Hamltonian, or to make available a larger range of gauge choices, or to go beyond perturbation theory.  Going back to Eq.~(1), and following the same reasoning$^{11}$ that leads from the operator formalism to the path-integral formalism in the calculation of S-matrix elements, we see that the vacuum expectation value of any product $Q(t)$ of $\delta\phi$s and $\delta\pi$s at the same time $t$ (now taking $t_0=-\infty$) is 
\begin{eqnarray}
\left\langle Q(t)\right\rangle&=&\int\prod_{{\bf x},t',a}d\delta\phi_{La}({\bf x},t')\;\prod_{{\bf x},t',a}\frac{d\delta\pi_{La}({\bf x},t')}
{2\pi}\prod_{{\bf x},t',a}d\delta\phi_{Ra}({\bf x},t')\;\prod_{{\bf x},t',a}\frac{d\delta\pi_{Ra}({\bf x},t')}{2\pi}\nonumber\\ &&
\times \exp\left\{-i\int_{-\infty}^t dt'\,\left[\sum_a \int d^{3}x\, \delta\dot{\phi}_{La}({\bf x},t')\delta\pi_{La}({\bf x},t')-\tilde{H}[\delta\phi_L(t'),\,\delta\pi_L(t');\,t']\right]\right\}\nonumber\\ && 
\times \exp\left\{i\int_{-\infty}^t dt'\,\left[\sum_a\int d^{3}x\, \delta\dot{\phi}_{Ra}({\bf x},t')\delta\pi_{Ra}({\bf x},t')-\tilde{H}\Big(\delta\phi_R({\bf x},t'),\,\delta\pi_R({\bf x},t');\,t'\Big)\right]\right\}
\nonumber\\&&\times \prod_{{\bf x},a} \delta\Big(\delta\phi_{La}({\bf x},t)-
\delta\phi_{Ra}({\bf x},t)\Big)\,\delta\Big(\delta\pi_{La}({\bf x},t)-\delta\pi_{Ra}({\bf x},t)\Big)\;Q\Big[\delta\phi_L(t),\delta\pi_L(t)\Big]\nonumber\\&&\times \Psi^*_0\Big[\delta\phi_L(-\infty)\Big]\,\Psi_0\Big[\delta\phi_R(-\infty)\Big]
\;.
\end{eqnarray}
Here the functional $\Psi_0[\delta\phi]$ is the wave function of the vacuum,$^{12}$
\begin{eqnarray}
&&\Psi_0[\phi(-\infty )]\propto  \exp\left(-\frac{1}{2}\sum_{a,b}\int d^{3}x\int d^{3}y \,{\cal E}_{ab}({\bf x},{\bf y})
\delta\phi_a({\bf x},-\infty)\,\delta\phi_b({\bf y},-\infty)\right)\nonumber\\&&~~~~=
\exp\left(-\frac{\epsilon}{2}\int_{-\infty}^t dt'\;e^{\epsilon t'}\sum_{a,b}\int d^{3}x\int d^{3}y\,{\cal E}_{ab}({\bf x},{\bf y})\,
\delta\phi_a(t')\,\delta\phi_b(t')\right)\;,
\end{eqnarray}
where ${\cal E}_{ab}$ is a positive-definite kernel.  For instance, for a real scalar field of mass $m$, 
\begin{equation}
{\cal E}({\bf x},{\bf y})\equiv \frac{1}{(2\pi)^{3}}\int d^{3}p\;e^{i{\bf p}\cdot({\bf x}-{\bf y})}\,
\sqrt{{\bf p}^2+m^2}\;.
\end{equation} 

As is well known, if the Hamiltonian is quadratic in the canonical conjugates $\delta\pi_a$ with a field-independent coefficient in the term of second order, then we can  integrate over the $\delta\pi_a$ by simply setting $\delta\dot{\phi}_a=\partial \tilde{H}/\partial\delta\pi_a$, and the quantity $\sum_a \delta\dot{\phi}_{a}(t')\delta\pi_{a}(t')-\tilde{H}\Big(\delta\phi(t'),\,\delta\pi(t');\,t'\Big)$ in Eq.~(A.24) then becomes the original Lagrangian.  We will not pursue this here, but will rather take up a puzzle that at first sight seems to throw doubt on the equivalence of the path integral formula (A.24), when we do {\em not} integrate out the $\pi$s, with the operator formalism.  

The puzzle is that, although the propagators for lines connecting left vertices to each other or right vertices to each other or left or right vertices to external lines are Greens functions of the sort that familiarly emerge from path integrals, what are we to make of the propagators arising from Eq.~(A.22) for lines connecting left vertices with right vertices?  These are not Greens functions; that is, they are solutions of {\em homogeneous} wave equations, not of inhomogeneous wave equations with a delta function source.  As we shall see, the source of these propagators lies in the delta functions in Eq.~(A.24).  It is these delta functions that tie together the integrals over the $L$ variables 
and over the $R$ variables, so that the expression (A.18)  does not factor into a product of these integrals.

In analyzing the consequences of Eq.~(A.24), it is convenient to condense our notation yet further, and let a variable $\xi_n(t)$ stand for all the $\delta\phi_a({\bf x},t)$ and $\delta\pi_a({\bf x},t)$, so that $n$ runs over positions in space and whatever discrete indices are used to distinguish different fields, plus a two-valued index that distinguishes $\delta\phi$ from $\delta\pi$.  With this understanding, Eq.~(A.24) reads
\begin{eqnarray}
&&\left\langle Q(t)\right\rangle=\int\prod_{t',n}\frac{d\xi_{Ln}(t')}{\sqrt{2\pi}}\prod_{t',n}\frac{d\xi_{Rn}(t')}{\sqrt{2\pi}}\nonumber\\&&~~~ \times\exp\left\{-i\int_{-\infty}^t dt'\,\tilde{L}\Big(\xi_L(t'),\dot{\xi}_L(t');\,t'\Big)\right\} \exp\left\{i\int_{-\infty}^t dt'\,\tilde{L}\Big[\xi_R(t'),\dot{\xi}_R(t');\,t'\Big]\right\}
\nonumber\\&&~~~\times \left(\prod_n \delta\Big(\xi_{Ln}(t)-\xi_{Rn}(t)\Big)\right)\;Q\Big(\xi_L(t)\Big) \Psi^*_0\Big(\xi_L(-\infty)\Big)\,\Psi_0\Big(\xi_R(-\infty)\Big)
\;,
\end{eqnarray}
where 
\begin{equation}
\tilde{L}[\xi(t'),\,\dot{\xi}(t');\,t']\equiv\sum_a\int d^{3}x\;\delta\pi_a({\bf x},t')\,\delta\dot{\phi}_a({\bf x},t')-\tilde{H}\Big[\delta\phi(t'),\,\delta\pi(t');\,t'\Big]\;.
\end{equation}
To expand in powers of the interaction, we split $\tilde{L}$ into a term $\tilde{L}_0$ that is quadratic in the fluctuations, plus an interaction term $-\tilde{H}_I$:
\begin{equation}
\tilde{L}=\tilde{L}_0-\tilde{H}_I\;,
\end{equation}
where
\begin{equation}
\tilde{L}_0[\xi(t'),\,\dot{\xi}(t');\,t']=\sum_a\int d^{3}x\delta\dot{\phi}_a({\bf x},t')\delta\pi_a({\bf x},t')-\tilde{H}_0\Big(\delta\phi(t'),\,\delta\pi(t');\,t'\Big)\;.
\end{equation}
As in calculations of the S-matrix, we will include the argument of the exponential in the vacuum wave functions along with the quadratic part of the Lagrangian, writing
\begin{eqnarray} 
&&\int^t_{-\infty}dt'\,\Bigg\{\tilde{L}_0[\xi_R(t'),\,\dot{\xi}_R(t');\,t']\nonumber\\&&~~~~~~~~~~+ \frac{i\epsilon}{2}\sum_{ab}\int d^{3}x\int d^{3}y\, {\cal E}_{ab}({\bf x},{\bf y})\, \delta\phi_{Ra}({\bf x},t')\,\delta\phi_{Rb}({\bf y},t')\Bigg\}\nonumber\\&&~~~~~~~\equiv
\frac{1}{2}\sum_{nn'}\sum_{t',t''}{\cal D}_{nt',mt''}^R \,\xi_{Rn}(t')\,\xi_{Rn'}(t'')\;,\\
&&\int_{\-\infty}^t dt'\Bigg\{\tilde{L}_0[\xi_L(t'),\,\dot{\xi}_L(t');\,t']\nonumber\\&&~~~~~~~~~~- \frac{i\epsilon}{2}\sum_{ab}\int d^{3}x\int d^{3}y\,{\cal E}_{ab}({\bf x},{\bf y})\,\delta\phi_{La}({\bf x},t')\,\delta\phi_{Lb}({\bf y},t')\Bigg\}\nonumber\\&&~~~~~~~\equiv
\frac{1}{2}\sum_{nn'}\sum_{t',t''}{\cal D}_{nt',n't''}^L \,\xi_{Ln}(t')\,\xi_{Ln'}(t'')\;
\end{eqnarray}
The vacuum wave function is the same for $\xi_L$ and $\xi_R$, but it is combined here with an exponential $\exp(-i\int \tilde{L}_0)$ for the $\xi_{Ln}$ and an exponential $\exp(+i\int \tilde{L}_0)$ for the $\xi_{Rn}$, which accounts for the different signs of the $i\epsilon$ terms in Eqs.~(A.31) and (A.32).  (The factor $e^{\epsilon t'}$ in Eq.~(A.25) is effectively equal to one for any finite $t'$, and has therefore been dropped.)  
We also express the product of delta functions in Eq.~(A.27) as a Gaussian:
\begin{eqnarray}
&&\prod_n \delta\Big(\xi_{Ln}(t)-\xi_{Rn}(t)\Big)\propto \exp\left(-\frac{1}{\epsilon'}\sum_n\Big(\xi_{Ln}(t)-\xi_{Rn}(t)\Big)^2\right)\nonumber\\&&~~~=
\exp\left(-\sum_{nn'}\sum_{t't''}{\cal C}_{nt',n't''}\Big(\xi_{Ln}(t')-\xi_{Rn}(t')\Big)
\Big(\xi_{Ln'}(t'')-\xi_{Rn'}(t'')\Big)\right)\;,~~~~
\end{eqnarray}
where
\begin{equation}
{\cal C}_{nt',n't''}\equiv \frac{1}{\epsilon'}\delta_{nn'}\,\delta(t'- t)\,\delta(t''- t)\;,
\end{equation}
and $\epsilon'$ is another positive infinitesimal.  

Following the usual rules for integrating a Gaussian times a polynomial, the integral is given by a sum over diagrams as described  above, but with a line that connects right vertices with each other (or with external lines) 
contributing a factor $-i\Delta^{RR}_{nt',n't''}$, a line that connects left vertices with each other (or with external lines) 
contributing a factor $i\Delta^{LL}_{nt',n't''}$, and a line that connects a right vertex where it is associated with $\xi_n(t')$ with a left vertex associated with $\xi_{n'}(t'')$ 
contributing a factor $i\Delta^{RL}_{nt',n't''}$, with the $\Delta$s determined by the
condition 
\begin{equation}
\left(\begin{array}{cc}{i\cal D}^R-{\cal C} & {\cal C} 
\\ {\cal C} & -i{\cal D}^L-{\cal C}\end{array}\right)\,\left(\begin{array}{cc}
-i\Delta^{RR} & i\Delta^{RL} \\ i(\Delta^{RL})^{\rm T} & i\Delta^{LL}\end{array} \right)=\left(\begin{array}{cc}
1 & 0 \\ 0 & 1 \end{array}\right)\;.
\end{equation}
This must hold whatever tiny value we give to $\epsilon'$, and so
\begin{equation}
{\cal D}^R\Delta^{RR}=1\;,~~~~~~~{\cal D}^L\Delta^{LL}=1\;,
\end{equation}
\begin{equation}
{\cal D}^R\Delta^{RL}=0\;,~~~~~~~{\cal D}^L\Big(\Delta^{RL}\Big)^{\rm T}=0\;,
\end{equation}
\begin{equation}
{\cal C}\Delta^{LL}={\cal C}\Delta^{RL}\;,~~~~~~~{\cal C}\Delta^{RR}=-{\cal C}(\Delta^{RL})^{\rm T}\;.
\end{equation}

The first Eq.~(A.36) is the usual inhomogeneous wave equation for the propagator, whose solution as well known is 
\begin{equation}
-i\Delta^{RR}_{nt',n't''}=\langle T\{\xi_n(t')\,\xi_{n'}(t'')\}\rangle\;,
\end{equation}
with the time-ordering dictated by the $+i\epsilon$ in Eq.~(A.31).  The second Eq.~(A.36) 
 is the complex conjugate of the first wave equation, whose solution is the complex conjugate of Eq.~(A.39):
\begin{equation}
i\Delta^{LL}_{nt',n't''}=\langle \bar{T}\{\xi_n(t')\,\xi_{n'}(t'')\}\rangle\;.
\end{equation}
Eqs.~(A.39) and (A.40) thus give the same propagators for lines connecting right vertices with each other or with external lines, and for lines connecting left vertices with each other or with external lines, as we we encountered in the operator formalism.  Equations (A.37) tell us that $\Delta^{RL}$ and $(\Delta^{RL})^{\rm T}$ satisfy the homogeneous versions of the wave equations satisfied by $\Delta^{RR}$ and $\Delta^{LL}$, but to find $\Delta^{RL}$ we also need an initial condition.  This is provided by the first of Eqs.~(A.38), which in more detail reads
\begin{equation}
i\Delta^{RL}_{nm}(t,t_2)=i\Delta^{LL}_{nm}(t,t_2)=\langle \bar{T}\{\xi_n(t)\xi_m(t_2)\}\rangle=\langle \xi_m(t_2)\xi_n(t)\rangle\;,
\end{equation}
in which we have used the fact that $t>t_2$.  This, together with the first of Eqs.~(A.37), tells us that 
\begin{equation}
i\Delta^{RL}_{nm}(t_1,t_2)=\langle \xi_m(t_2)\xi_n(t_1)\rangle\;,
\end{equation}
which is the same propagator for internal lines connecting right vertices with left vertices that we found in the operator formalism.

\vspace{9pt}

\noindent
{\bf 5. Tree Graphs and Classical Solutions}.

We will now verify the remark made in Section I, that the usual approach to the calculation of non-Gaussian correlations, of solving the classical field equations beyond the linear approximation, simply corresponds to the calculation of tree diagrams in the ``in-in'' formalism.  This is a well-known result$^{13}$ in the usual applications of quantum field theory, but  some modifications in the usual argument are needed in the ``in-in'' formalism, in which the vacuum persistence functional is always unity whether or not we add a current term to the Lagrangian.  

We begin by introducing a generating functional $W[j,t,g]$ for correlation functions of fields at a fixed time $t$:
\begin{equation}
e^{W[J,t,g]/g}\equiv
 \Bigg\langle {\rm vac, in}\Bigg|e^{\;\frac{1}{g}\sum_a\int d^{3}x\,\delta\phi_a({\bf x},t)J_a({\bf x})}\Bigg|{\rm vac, in}\Bigg\rangle_g\;,
\end{equation}
where $J_a$ is an arbitrary current, and $g$ a real parameter,
with the subscript $g$ indicating that the expectation value is to be calculated using a Lagrangian density multiplied with a factor $1/g$.  (This is different from the usual definition of the effective action, because here we are not introducing the current into the Lagrangian.)
The quantity of physical interest is of course $W[J,t,1]$, from which expectation values of all products of fields can be found by expanding in powers of the current.

Using Eq.~(A.27), we can calculate $W$ as the path integral
\begin{eqnarray}
&& e^{W[J,t,g]/g}=\int \prod \delta\phi_L\;\int\prod 
\delta\pi_L\;\int\prod \delta\phi_R\;\int\prod \delta\pi_R 
\nonumber\\&&\times\,\exp\left(-i\int_{-\infty}^t 
dt'\;\frac{1}{g}\tilde{L}[\delta\phi_L,\delta\pi_L;t']\right)\nonumber\\
&& \times\,\exp\left(+i\int_{-\infty}^t 
dt'\;\frac{1}{g}\tilde{L}[\delta\phi_R,\delta\pi_R;t']\right)\nonumber\\&&~~~\times\,\prod\delta[\phi_L(t)-\delta\phi_R(t)]\;\prod\delta[\delta\pi_L(t)-\delta\pi_R(t)] \nonumber\\&&~~~\times\,
e^{\;\frac{1}{g}\sum_a\int d^{3}x\,\delta\phi_a({\bf x},t)J_a({\bf x})}\cdots\nonumber\\&&~~~\times\,
\Psi_{\rm vac}[\delta\phi_L(-\infty)]\;\Psi_{\rm vac}[\delta\phi_R(-\infty)]
\end{eqnarray}
The usual power-counting arguments$^{13}$ show that the $L$ loop contribution to 
$W[J,t,g]$ has a $g$-dependence given by a factor $g^{-L}$.  
For  $g\rightarrow 0$, $W$ is thus given by the sum of all {\em tree} graphs.  The integrals over $\delta\phi_L$, $\delta\pi_L$, $\delta\phi_L$, $\delta\pi_L$ are dominated in the limit $g\rightarrow 0$ by fields where $\tilde{L}$ is stationary, i.e., where 
$$\delta\phi_L=\delta\phi_R=\delta\phi^{\rm classical}$$
$$\delta\pi_L=\delta\pi_R=\delta\pi^{\rm classical}$$
with  $\delta\phi^{\rm classical}$ and $\delta\pi^{\rm classical}$ the solutions of the classical field equations with the initial conditions that the fields go to free fields such as (14)--(16) satisfying the initial conditions (20) at $t\rightarrow -\infty$.  Since the $L$ and $R$ fields take the same values at this stationary point, the action integrals cancel, and we conclude that  
\begin{equation}
\Big[W[J,t,1]\Big]_{\rm zero\;loops}=\sum_a\int d^{3}x\;\delta\phi^{\rm classical}_a({\bf x},t)\,J_a({\bf x})\;.
\end{equation}
Expanding in powers of the current, this shows that in the tree approximation the expectation value of any product of fields is to be calculated by taking the product of the fields obtained by solving the non-linear classical field equations with suitable free-field initial conditions, as was to be proved.

\pagebreak

\begin{center}
{\bf REFERENCES}
\end{center}
\nopagebreak

\begin{enumerate}

\item For a review, see N. Bartolo, E. Komatsu, S. Matarrese, and A. Riotto, astro-ph/0406398.

\item J. Maldacena, JHEP {\bf 0305}, 013 (2003) (astro-ph/0210603).  For other work on this problem, see A. Gangui, F. Lucchin, S. Matarrese,and S. Mollerach, Astrophys. J. {\bf 430}, 447 (1994) (astro-ph/9312033); P. Creminelli, astro-ph/0306122; P. Creminelli and M. Zaldarriaga, astro-ph/0407059; G. I. Rigopoulos, E.P.S. Shellard, and B.J.W. van Tent, astro-ph/0410486; and ref. 3.

\item J. Schwinger, Proc. Nat. Acad. Sci. US {\bf 46}, 1401 (1961).  Also see L. V. Keldysh, Soviet Physics JETP {\bf 20}, 1018 (1965); B. DeWitt, {\em The Global Approach to Quantum Field Theory} (Clarendon Press, Oxford, 2003): Sec. 31.  This formalism has been applied to cosmology by E. Calzetta and B. L. Hu, Phys. Rev. D {\bf 35}, 495 (1987); M. Morikawa, Prog. Theor. Phys. {\bf 93}, 685 (1995); N. C. Tsamis and R. Woodard, Ann. Phys. {\bf 238}, 1 (1995); {\bf 253}, 1 (1997); N. C. Tsamis and R. Woodard, Phys. Lett. {\bf B426}, 21 (1998); V. K. Onemli and R. P. Woodard, Class. Quant. Grav. {\bf 19}, 407 (2002); T. Prokopec, O. Tornkvist, and R. P. Woodard, Ann. Phys. {\bf 303}, 251 (2003); T. Prokopec and R. P. Woodard, JHEP {\bf 0310}, 059 (2003); T. Brunier, V.K. Onemli, and R. P. Woodard, Class. Quant. Grav. {\bf 22}, 59 (2005), but not (as far as I know) to the problem of calculating cosmological correlation functions.

\item F. Bernardeau, T. Brunier, and J-P. Uzam, Phys. Rev. D {\bf 69}, 063520 (2004).

\item   The constancy of this quantity outside the horizon has been used in 
various special
cases by J. M. Bardeen, Phys. Rev. {\bf D22}, 1882 (1980);  
D. H. Lyth, Phys. Rev. {\bf D31}, 1792 (1985).  For reviews, see 
J. Bardeen, in {\em Cosmology and Particle Physics}, eds. Li-zhi Fang 
and A. Zee (Gordon \& Breach, New York, 1988); A. R. Liddle and D. H. 
Lyth, {\em Cosmological Inflation and Large Scale Structure} 
(Cambridge 
University Press, Cambridge, UK, 2000).

\item R. S. Arnowitt, S. Deser, and C. W. Misner, in {\em Gravitation: An Introduction to Current Research}, ed. L. Witten (Wiley, New York, 1962): 227.  This classic article is now available as gr-qc/0405109.

\item V. S. Mukhanov, H. A. Feldman, and R. H. Brandenberger, Physics Reports {\bf 215}, 203 (1992); E. D. Stewart and D. H. Lyth, Phys. Lett. B {\bf 302}, 171 (1993).

\item N. C. Tsamis and R. Woodard, ref. 3.

\item A. Vilenkin and L. H. Ford, Phys. Rev. D. {\bf 26}, 1231 (1982); A. Vilenkin, Nucl. Phys. {\bf B226}, 527 (1983).

\item R. Kubo, J. Math. Phys. {\bf 4}, 174 (1963).

\item See, e.g., S. Weinberg, {\em The Quantum Theory of Fields -- Volume I} (Cambridge, 1995): Sec. 9.1.

\item {\em ibid.}, Sec. 9.2.

\item S. Coleman, in {\em Aspects of Symmetry} (Cambridge University Press, Cambridge, 1985): pp 139--142.

\end{enumerate}
\end{document}